\documentclass[prb,amssymb,twocolumn,reprint,superscriptaddress,showpacs,floatfix]{revtex4}
\usepackage{bm}
\usepackage{graphicx}
\usepackage{color}
\usepackage{amsmath}

\begin{document}

\title{Thermoelectric transport through strongly correlated quantum dots}

\author{T. A. Costi}
\affiliation{Institut f\"ur Festk\"orperforschung, 
Forschungszentrum J\"ulich, 52425 J\"ulich, Germany}
\author{V. Zlati\'{c}}
\affiliation{Institut f\"ur Festk\"orperforschung, Forschungszentrum
J\"ulich, 52425 J\"ulich, Germany} 
\affiliation{Institute of Physics, 10001 Zagreb, Croatia}

\date{\today}
\begin{abstract}
The thermoelectric properties of strongly correlated quantum dots, described
by a single level Anderson model coupled to conduction electron leads, is 
investigated using Wilson's numerical renormalization 
group method. We calculate the electronic contribution, $K_{\rm e}$, to the 
thermal conductance, the thermopower, $S$, and the electrical conductance, $G$, 
of a quantum dot as a function of both temperature, $T$, and gate voltage, ${\rm v}_g$, 
for strong, intermediate and weak Coulomb correlations, $U$, on the dot. For strong 
correlations and in the Kondo regime, we find that the thermopower exhibits two sign changes, 
at temperatures $T_{1}({\rm v}_g)$ and $T_{2}({\rm v}_g)$ with $T_{1}< T_{2}$. We find that 
$T_{1}> T_{p}({\rm v}_g)\approx T_{K}({\rm v}_g)$, where $T_{p}({\rm v}_g)$ is the position of the 
Kondo induced peak in the thermopower, 
$T_{K}({\rm v}_g)$ is the Kondo scale, and $T_{2}= O(\Gamma)$, where $\Gamma$ is the level width. 
The loci of  $T_{1}({\rm v}_g)$ and $T_{2}({\rm v}_g)$ merge at a critical
gate voltage ${\rm v}_g={\rm v}_g^{c}(U/\Gamma)$, beyond which no sign change occurs at finite 
gate voltage (measured relative to mid-valley).
We determine ${\rm v}_g^{c}$ for different $U/\Gamma$ finding that ${\rm v}_g^{c}$
coincides, in each case, with entry into the mixed valence regime. No sign change is 
found outside the Kondo regime, or, for weak correlations $U/\Gamma \lesssim 1$, 
making such a sign change in $S(T)$ a particularly sensitive signature 
of strong correlations and Kondo physics. The relevance of this to recent thermopower
measurements of Kondo correlated quantum dots is discussed. The results for quantum 
dots are compared also 
to the relevant transport coefficients of dilute magnetic impurities 
in non-magnetic metals: the electronic contribution, $\kappa_{\rm e}$, to the
thermal conductivity, the thermopower, $S$, and the impurity contribution to the 
electrical resistivity, $\rho$.
In the mixed valence and empty orbital regimes, we find, as a function of
temperature, two peaks in $K_{\rm e}$ as compared to a single peak in $\kappa_{\rm e}$, and
similarly, $G(T)$ exhibits a finite temperature peak on entering the mixed
valence regime, whereas such a pronounced peak is absent in $\rho(T)$ even far into
the empty orbital regime. We compare and contrast the  figure of merit,
power factor and the extent of violation of the Wiedemann-Franz law in quantum
dots and dilute magnetic impurities. The extent of temperature scaling in the 
thermopower and thermal conductance of quantum dots in the Kondo regime is 
discussed. 
\end{abstract}

\maketitle

\section{Introduction}
 
Materials with potentially useful thermoelectric properties are 
currently under intense theoretical and experimental investigation, mainly due to
the prospect of applications, e.g., for conversion of waste heat into electricity 
in thermoelectric generators, for applications to refrigeration, 
or for on-chip cooling and energy efficiency in microelectronics 
applications\cite{mahan.98,kanatzidis.10,terasaki.97,arita.08,
lackner.06,paschen.06,bentien.07,sales.96,matusiak.09,
hsu.04,venkatasubramanian.01,cai.08,harman.96,beyer.02}. 
Apart from possible applications, thermoelectric materials can also serve as 
an interesting testing ground for theoretical approaches to electrical and 
thermal transport in solids\cite{zlatic.94,costi.94,nca,zlatic.05,grenzenbach.09}. 
As the scale of the individual components in 
semiconducting devices  is approaching  the nano-size, a description of  
thermal transport through quantum dots is also attracting a lot of experimental 
and theoretical attention\cite{scheibner.05,kim.02,dong.02}.  

In this paper we address the thermoelectric properties of
a nanoscale size quantum dot exhibiting the Kondo effect, which we 
describe in terms of a single level Anderson impurity model with two conduction
electron leads at fixed chemical potentials. The quantum dots that we consider have 
sizes of $10-100{\rm\, nm}$ and can be tuned from the Kondo to the mixed 
valence and empty orbital regimes by a gate voltage
\cite{goldhaber-gordon.98,schmid.98,cronenwett.98,vanderwiel.00}. 
Short segments of carbon nanotubes \cite{nygard.00}
connected to leads exhibit similar physics, so our results could also be of 
relevance to such systems. Very recent experiments on nanoscale quantum dots 
\cite{scheibner.05,scheibner.07} are beginning to probe the effect of Kondo correlations on 
the thermopower, although as we shall argue in the conclusions 
a quantitative comparison with theory is still some way off. 
The thermoelectric
properties of dilute magnetic impurities in non-magnetic metals, such as 
Ce$_{x}$La$_{1-x}$Al$_{3}$ and  Ce$_{x}$La$_{1-x}$B$_{6}$, are closely related
to those of quantum dots (see Sec.~\ref{model}) so we discuss these here also. 
Understanding the thermoelectric properties of magnetic impurities is also a useful 
starting point for understanding those of heavy fermions within the 
dynamical mean field theory approach\cite{grenzenbach.09} although, in these systems,
crystal field effects and non-resonant channels play a crucial role for the thermopower, 
and need to be taken into account for a quantitative comparison to experiment 
\cite{zlatic.05,zlatic.94,nca}.

The approach that we use in this paper, Wilson's numerical renormalization group (NRG) method
\cite{wilson.75,kww.80,bulla.08}, gives reliable results for transport properties in all parameter and 
temperature regimes of interest\cite{costi.94}. The present calculations
were carried out for the Anderson model with finite Coulomb repulsion, as is appropriate
for nanoscale size quantum dots. We implemented recent developments in the calculation of 
dynamical quantities within the NRG, including the use of the self-energy\cite{bulla.98} 
and the full density matrix (FDM) generalization \cite{fdm.07} (see also  
Ref.~\onlinecite{peters.06,toth.08}) of the reduced 
density matrix approach\cite{hofstetter.00} within the complete basis set of 
eliminated states\cite{anders.05}. In particular, the FDM approach
allows calculations of dynamical properties at all excitation energies $\omega$ 
relative to the temperature $T$, thereby simplifying the calculation of 
transport properties which require knowledge of excitations, $\omega$, above and
below the temperature \cite{costi.94}.

The outline of the paper is as follows. In Sec.~\ref{model}-\ref{transport} we describe
the Anderson impurity model for quantum dots and dilute magnetic impurities
and we specify the relevant transport quantities that we calculate for these
two different physical realizations of the model. The NRG method used in this
paper is described in Sec.~\ref{method} together with results for occupancies
which we use to define Kondo, mixed valence and empty orbital regimes in the
strong correlation limit. Sec.~\ref{Tdependence} presents the temperature
dependent transport properties of quantum dots and Sec.~\ref{Tdependence-compare}
compares these to the corresponding quantities for dilute magnetic impurities.
Results for the  figure of merit, power factor and Lorenz number ratios 
for quantum dot and magnetic impurity systems are presented in Sec.~\ref{zt}. 
Sec.~\ref{scaling} investigates the extent to which universal 
scaling functions apply to the thermopower and thermal conductance of 
quantum dots in the Kondo regime. 
In Sec.~\ref{gate-dependence} we present our results for the
gate voltage (local level) dependence of transport quantities for quantum dots 
(magnetic impurities). Conclusions and a discussion of the relevance of our results to
recent experiments on nanoscale size quantum dots is presented in Sec.~\ref{conclusions}.
Appendix~\ref{reduction} discusses the reduction of the two-lead Anderson model to a
single channel model, Appendix~\ref{extra-results} contains some additional results for moderately 
and weakly correlated quantum dots, and Appendix~\ref{full-density-matrix} provides details of the
FDM approach \cite{fdm.07} and an alternative detailed derivation of the 
FDM expression for local Green's functions, which we have used to obtain
the results in this paper. Finally Appendix~\ref{transport-derivations}
gives an outline of the derivation of thermopower and thermal conductance for 
quantum dots.

\section{Model}
\label{model}
A nanoscale quantum dot is described by the single level Anderson impurity
model with two conduction electron leads
\begin{eqnarray}
H &=& \sum_{\alpha k\sigma}\epsilon_{\alpha k\sigma}c_{\alpha k\sigma}^{\dagger}c_{\alpha k\sigma} 
+ \sum_{\sigma}\varepsilon_{d}\,d_{\sigma}^{\dagger}d_{\sigma} + 
U n_{d\uparrow}
n_{d\downarrow}\nonumber\\
& + &\sum_{\alpha k\sigma}t_{\alpha}(c_{\alpha k\sigma}^{\dagger}d_{\sigma}+h.c.).
\label{qdot-two-leads}
\end{eqnarray}
Here, $\epsilon_{\alpha k\sigma}$ is the kinetic energy of conduction 
electrons with wavenumber $k$ and spin $\sigma$ in lead $\alpha=(L,R)$, 
$\varepsilon_{d}$ is the local level energy, $U$ is the Coulomb repulsion 
on the dot and $t_{\alpha}$ is the tunnel matrix element
of the dot level to conduction electron states in lead $\alpha=(L,R)$. The operators 
$c^{\dagger}_{\alpha k\sigma} (c_{\alpha k\sigma})$ 
create (destroy) conduction electron states $|\alpha k\sigma\rangle$ 
and $d_{\sigma}^{\dagger} (d_{\sigma})$ create (destroy) local d-level 
states $|\sigma\rangle$. We assume a flat density of
states of magnitude $N_{F}=1/2D$ per spin channel for both leads, where $D=1$ is the 
half bandwidth of each lead. The single-particle broadening (half-width at half-maximum) 
of the d-level is given
by $\tilde{\Gamma}=\tilde{\Gamma}_{L}+\tilde{\Gamma}_{R}$, where $\tilde{\Gamma}_{L,R}=\pi N_{F}t_{L,R}^{2}$ 
are the contributions to the broadening from the left and right leads. In this paper, we follow
the convention used in quantum dot work and use as unit of energy not $\tilde{\Gamma}$, but 
the full-width at half-maximum $\Gamma=2\tilde{\Gamma}$.

Since the $d$-state of the quantum dot in (\ref{qdot-two-leads}) 
only couples to the even combination $a_{ek\sigma}\sim t_{L}c_{Lk\sigma}+t_{R}c_{Rk\sigma}$
of the lead electron states, one can show 
(see  Appendix~\ref{reduction})
that, to a very good approximation, the above model can be reduced to the following 
single-channel Anderson model 
\begin{eqnarray}
H &=& \sum_{k\sigma}\epsilon_{ek\sigma}a_{ek\sigma}^{\dagger}a_{ek\sigma} 
+ \sum_{\sigma}\varepsilon_{d}\,d_{\sigma}^{\dagger}d_{\sigma} + 
U n_{d\uparrow}n_{d\downarrow}\nonumber\\
& + &t\sum_{k\sigma}(a_{ek\sigma}^{\dagger}d_{\sigma}+h.c.),
\label{siam}
\end{eqnarray}
where the tunneling amplitude $t$ is given by $t^{2}=t_{L}^{2}+t_{R}^{2}$,
so that the hybridization strength of the dot to the leads is given by
$\tilde{\Gamma}=\tilde{\Gamma}_{L}+\tilde{\Gamma}_{R}$. This is also the appropriate
model for describing dilute magnetic impurities in non-magnetic metals\cite{costi.94}. 
In fact, for both systems (see below and Appendix~\ref{reduction}) the calculation 
of the linear transport properties reduces to the calculation of the 
equilibrium d-level spectral density of the single-channel model
\begin{equation}
A(\omega)=-\frac{1}{\pi}{\rm Im}[G_{d\sigma}(\omega+i\delta)],
\end{equation}
where 
$G_{d\sigma}(\omega+i\delta)=\langle\langle d_{\sigma}; 
d_{\sigma}^{\dagger}\rangle\rangle$
is the Fourier transform of the retarded d-level Green function of (\ref{siam}).
Hence, all results in this paper, including those for dilute magnetic 
impurities, are obtained by solving the single-channel model (\ref{siam}) 
using the NRG (as explained in Sec.~\ref{method}) to obtain $A(\omega,T)$. 
\section{Transport quantities}
\label{transport}
\subsection{Quantum dots}
Thermoelectric transport through the quantum dot (\ref{qdot-two-leads}) is calculated 
for a steady state situation in which a small external bias voltage, 
$\delta V=V_{L}-V_{R}$, and a small temperature gradient $\delta T$
is applied between the left and right leads. Left and right leads are
then at different chemical potentials $\mu_{L}$ and $\mu_{R}$, and 
temperatures $T_{L}$ and $T_{R}$, 
with $e\delta V=\mu_{L}-\mu_{R}$ and $\delta T=T_{L}-T_{R}$. We follow the 
approach for deriving the electrical conductance, $G(T)$, the thermal 
conductance, $K_{\rm e}(T)$, and thermoelectric power, $S(T)$,
through an interacting quantum dot\cite{meir.93,hershfield.91,jauho.94} using 
the non-equilibrium Green's function formalism. For completenes, an outline of 
this derivation \cite{kim.02,dong.02} can be found in Appendix~\ref{transport-derivations}. 
The final expressions are given by
\begin{eqnarray}
 G(T) &=& e^{2}I_{0}(T)\label{transport-expressions-G}\\
S(T) &=& -\frac{1}{|e|T}\frac{I_{1}(T)}{I_{0}(T)}\label{transport-expressions-S}\\
K_{\rm e}(T) &=& \frac{1}{T}\left[I_{2}(T) - \frac{I_{1}^{2}(T)}{I_{0}(T)} \right]
\label{transport-expressions-K}
\end{eqnarray}
where $I_{n},n=0,1,2$ are the transport integrals
\begin{equation}
I_{n}(T) = \frac{2}{h}\int d\omega\; \omega^{n}{\cal T}(\omega)(-\frac{\partial f}{\partial \omega}).
\label{trans-dot}
\end{equation}
Here, $e$ denotes the magnitude of the electronic charge and $h$ denotes Planck's constant.
The quantity ${\cal T}(\omega)$ is related to the spectral density $A(\omega)$ via
\begin{eqnarray}
{\cal T}(\omega) 
&=& 4\pi\frac{\tilde{\Gamma}_{L}\tilde{\Gamma}_{R}}{\tilde{\Gamma}_{L}+\tilde{\Gamma}_{R}}A(\omega).
\end{eqnarray}
At $T=0$, the conductance acquires the value 
\begin{eqnarray}
G(0) &=& \frac{2e^{2}}{h}{\cal T}(0)\\
&=&\frac{2e^{2}}{h}\frac{4\tilde{\Gamma}_{L}\tilde{\Gamma}_{R}}
{(\tilde{\Gamma}_{L}+\tilde{\Gamma}_{R})^{2}}\sin^{2}(\pi n_{d}/2)
\end{eqnarray}
where $n_{d}$ is the occupancy of the dot and we have used the Friedel sum rule, 
$$
A(\omega=0,T=0)=\frac{1}{\pi(\tilde{\Gamma}_{L}+\tilde{\Gamma}_{R})}\sin^{2}(\pi n_{d}/2).
$$ 
For integer occupation, $n_{d}=1$, and equal coupling to the leads, 
$\tilde{\Gamma}_{L}=\tilde{\Gamma}_{R}$, the 
conductance reaches the unitary value $2e^{2}/h$, which we henceforth denote by $G_{0}$.
\subsection{Dilute magnetic impurities}
It is of interest to compare the transport properties of a quantum dot, with
the corresponding quantities for electrons scattering from a dilute concentration, 
$n_{i}\ll 1$, of magnetic impurities in a clean host metal with 
constant density of states $N_{F}$  per spin. 
As for quantum dots, the relevant model for such dilute magnetic impurities is the 
single channel Anderson model (\ref{siam}) with hybridization strength $\tilde{\Gamma}$. 
In order to obtain the thermopower, $S(T)$, the thermal conductivity, 
$\kappa_{\rm e}(T)$, and resistivity, $\rho(T)$ (or conductivity $\sigma=1/\rho$) for such a
dilute concentration of magnetic impurities we use the Kubo formalism, 
see Appendix~A of
Ref.~\onlinecite{costi.94} for the details, and find for these quantities 
\begin{eqnarray}
 \rho(T) &=& \frac{1}{e^{2}M_{0}(T)}\label{imp1}\\
S(T) &=& -\frac{1}{|e|T}\frac{M_{1}(T)}{M_{0}(T)}\label{imp2}\\
\kappa_{\rm e}(T) &=& \frac{1}{T}\left[M_{2}(T) - \frac{M_{1}^{2}(T)}{M_{0}(T)} \right].\label{imp3}
\end{eqnarray}
The transport integrals $M_{n},n=0,1,2$, appearing here, are now defined by
\begin{equation}
M_{n}(T) = \int d\omega\; \omega^{n}\tau(\omega,T)(-\frac{\partial f}{\partial \omega}),
\label{trans-imp}
\end{equation}
where $\tau(\omega,T)$ is the transport time of electrons, which is given in terms of the impurity
spectral density $A(\omega,T)$ by
\begin{equation}
\frac{1}{\tau(\omega,T)} =  \frac{n_{i}}{N_{F}}2\tilde{\Gamma} A(\omega,T).
\end{equation}
In order to compare impurity transport properties with those of quantum dots, using
the same units, we shall use rescaled quantities, e.g. for quantum dots 
$G/G_{0}$ and $K_{\rm e}/G_{0}$ and for impurities $\rho/\rho_{0}$ and $\kappa_{\rm e}\rho_{0}$ 
where 
\begin{equation}
\rho_{0}=2n_{i}/\pi N_{F}e^{2},\label{unitary-rho}
\end{equation} 
is the unitary resistivity of electrons 
scattering from a dilute concentration $n_{i}$ of magnetic impurities.

While the physics governing the transport properties of electrons scattering from dilute
magnetic impurities, described by (\ref{siam}), is expected to be similar to that
governing the transport properties of electrons through quantum dots (also described by (\ref{siam})), 
differences are also expected, particularly for the respective thermopowers or the thermal conductance 
(conductivity), for the following reason: the transport expressions for quantum dots arise from 
integrals $I_{n},n=0,1,2$ which involve the $n$'th moments of $A(\omega,T)$ convoluted with the
derivative of a Fermi function, whereas those for magnetic impurities arise from n'th moments 
of $1/A(\omega,T)$ convoluted with the same derivative. At low temperatures, 
a Sommerfeld expansion for $I_{1}$ and $M_{1}$ results in different signs 
for the thermopower in the two different situations, since derivatives of $A$ and $1/A$ 
have opposite signs. On the other hand, at higher temperatures, moments of $A$ and $1/A$ 
are determining factors for transport. Particularly the moments $I_{1}$ ($M_{1}$), 
entering the thermopower, and $I_{2}$ ($M_{2}$), entering the thermal conductance (conductivity), 
probe differences in the behavior or $A(\omega,T)$ and $1/A(\omega,T)$ at high temperature.
Consequently we expect significant quantitative differences for the thermopower and 
thermal conductance (conductivity) of quantum dots and dilute magnetic impurities 
at high temperatures. We discuss these differences in Sec.~{\ref{Tdependence-compare}}.

\section{NRG approach}
\label{method}
\subsection{NRG and dynamical quantities}
\begin{figure}[t]
\includegraphics[scale=0.3]{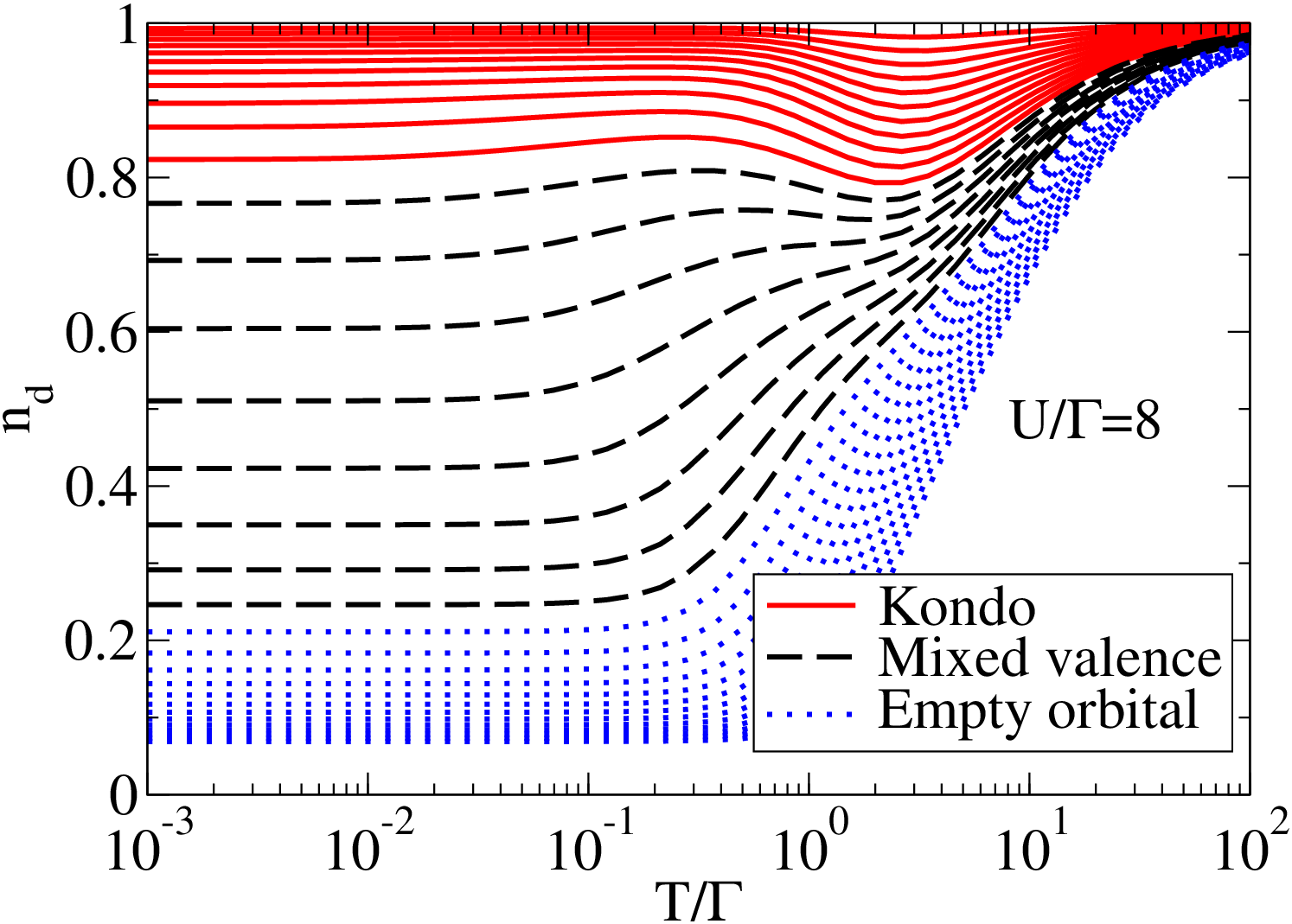}
\caption{  Temperature dependence of the occupancy for different values of
the  gate voltages ${\rm v}_g>0$ in the Kondo (solid lines), mixed valence (dashed lines)
and empty orbital (dotted lines) regimes. We define these regimes by 
$|n_{d}(T=0)-1| \lesssim 0.25$, $|n_{d}(T=0)-0.50| \lesssim 0.25$ and 
$|n_{d}(T=0)| \lesssim 0.25$, respectively. 
} \label{figure1}
\end{figure}
\vspace{1cm}
\begin{figure}[t]
\includegraphics[scale=0.3]{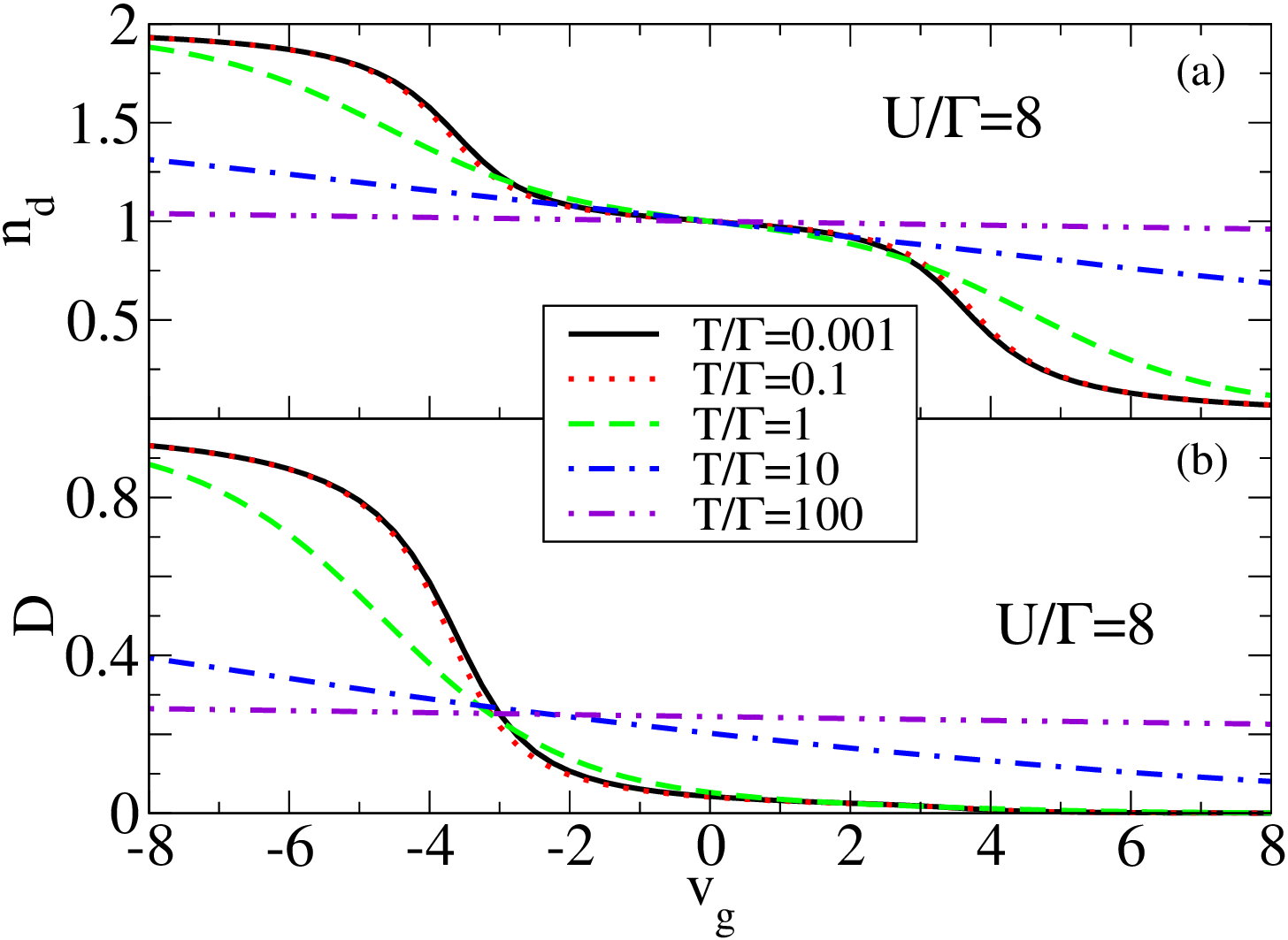}
\caption{  Dependence on  gate voltage, ${\rm v}_g=(\varepsilon_{d}+U/2)/\Gamma$, 
of, (a), the occupancy of the quantum dot, $n_{d}$ , and, (b), the double occupancy, 
$D=\langle n_{d\uparrow}n_{d\downarrow}\rangle$, at several temperatures. 
} \label{figure2}
\end{figure}

We calculate the spectral function $A(\omega,T)$ and the transport properties
of quantum dots, by using the NRG approach\cite{wilson.75,kww.80,bulla.08}. This
method is numerically exact and can be used to calculate both static thermodynamic
properties as well as finite temperature dynamic and transport properties 
\cite{costi.94}. In brief, the NRG procedure \cite{wilson.75,kww.80} consists 
of the following steps, (i), a logarithmic mesh of $\varepsilon_k^{n}=D\Lambda^{-n}$ 
is introduced about the Fermi level $\varepsilon_{F}=0$,
and, (ii), a unitary transformation of the $a_{ek\sigma}$ in (\ref{siam}) 
is performed such that $f_{0\sigma}=\sum_{k}a_{ek\sigma}$ is the first operator in a new 
basis, $f_{n\sigma},\,n=0,1,\dots$, which tridiagonalizes 
$H_{c}=\sum_{k\sigma}\epsilon_{ek\sigma}a_{ek\sigma}^{\dagger}a_{ek\sigma}$, 
i.e. $H_{c}\rightarrow \sum_{\sigma}\sum_{n=0}^{\infty}t_{n}
(f_{n+1\sigma}^{\dagger}f_{n\sigma}+ h.c.).$, where the hoppings 
$t_{n}\approx\Lambda^{-n/2}$ for a flat conduction band\cite{kww.80}.
The Hamiltonian (\ref{siam}) with the above discretized 
form of the kinetic energy is now iteratively diagonalized 
by defining a sequence of finite size Hamiltonians 
\begin{eqnarray}
H_{m} &=& \sum_{\sigma}\varepsilon_{d}\,d_{\sigma}^{\dagger}d_{\sigma} + 
U n_{d\uparrow}n_{d\downarrow}
+t\sum_{\sigma}(f_{0\sigma}^{\dagger}d_{\sigma}+h.c.)\nonumber\\
&+&\sum_{n=0,\sigma}^{m-1}t_{n}(f_{n+1\sigma}^{\dagger}f_{n\sigma}+ h.c.) 
\label{discrete-model}
\end{eqnarray}
for $m\ge 0$ up to a maximum chain length $N$. 
For each $m$, this yields the excitations $E_{p}^{m}$ and many 
body eigenstates $|pm\rangle$ of $H_m$ at a corresponding
set of energy scales $\omega_{m}$ defined by the smallest scale in $H_m$,
$\omega_{m}=t_{m}\approx\Lambda^{-\frac{m-1}{2}}$.  Since the
number of states grows as $4^m$, for $m>6$ only the lowest 600 or so states 
of $H_m$ are retained.  These are used as a basis for constructing $H_{m+1}$. 
For $m>6$, both the retained and eliminated (high energy) states of $H_{m}$, 
together with the corresponding eigenvalues, are stored. This information is
subsequently used to evaluate the spectral 
function $A(\omega,T)$ within the FDM approach \cite{fdm.07} 
described in Appendix~\ref{full-density-matrix}. This evaluation makes use of, (i),
the completeness of eliminated states \cite{anders.05}, allowing a multiple-shell
evaluation of Green's functions \cite{costi.97}, avoiding double counting of excitations, 
and, (ii), the reduced density matrix approach to Greens functions, 
introduced to the NRG by Hofstetter \cite{hofstetter.00}. In addition, 
we calculate the spectral function via the correlation part of the 
self-energy $\Sigma(\omega,T)$ following Bulla, Pruschke and Hewson \cite{bulla.98}, via
\begin{eqnarray}
A(\omega,T) &=& -\frac{1}{\pi}{\rm Im} \left[\frac{1}
{\omega-\epsilon_{d}+i\Gamma - \Sigma(\omega,T)}\right],\nonumber\\
\Sigma(\omega,T) &=& U \frac{\langle\langle n_{d,-\sigma}d_{\sigma};
d_{\sigma}^{\dagger}\rangle\rangle}{\langle\langle d_{\sigma};d_{\sigma}^{\dagger}\rangle\rangle}
\equiv U \frac{F_{\sigma}(\omega,T)}{G_{d\sigma}(\omega,T)}\label{correlationGF}
\end{eqnarray}

Since the FDM entering the definition of the Green's functions, see 
Appendix~\ref{full-density-matrix}, contains the complete spectrum from all NRG iterations, 
asymptotically high and low temperatures can be investigated more easily than within
previous approaches \cite{costi.94}, which involved at a given temperature $T$, choosing
an appropriate energy shell to extract $A(\omega,T)$. In addition, the regime $\omega\ll T$,
which was problematical in previous approaches, can now be addressed, since contributions 
from all excitations (for all energy shells) are taken into account in the expression for
the Green's function within the FDM approach.

\subsection{Calculations}
The calculations reported here have been carried out for a discretization parameter $\Lambda=1.75$,
retaining $660$ states per NRG iteration and a hybridization strength 
$\tilde{\Gamma} = 0.01$ (in units of the half-bandwidth $D=1$). The maximum chain length
diagonalized was $N=68$. We use the
full width $\Gamma=2\tilde{\Gamma}=0.02$ as our energy unit throughout. 
Results for a wide range of temperatures from $T/\Gamma \ll 1$ to $T/\Gamma \gg 1$ were obtained
to fully characterize the transport properties of quantum dots and dilute magnetic impurities. We note
that, in practice, the regime $T/\Gamma \gg 1$ is probably not accessible in experiment due to 
other effects which become important at high temperature, and which we do not take into account, e.g.
phonons, multiple levels, crystal field states etc. Calculations for strong ($U/\Gamma=8, 6$), 
moderate ($U/\Gamma=3$) and weak ($U/\Gamma=1$) correlations were carried out for
a range of  dimensionless gate voltages, ${\rm v}_g$, defined by
$$
{\rm v}_g = \frac{\varepsilon_{d}+U/2}{\Gamma} = 0.25n, n=\pm 1, \pm 2,\dots,\pm 32.
$$
With this definition, the gate voltage for mid-valley occurs at $v_{g}=0$ for all $U$.
Due to particle-hole symmetry, calculations were carried out for ${\rm v}_g>0$,
with those for ${\rm v}_g<0$ being obtained via a particle-hole transformation. This results
in ${\rm v}_g\rightarrow -{\rm v}_g$, occupancy $n_d\rightarrow 1-n_{d}$, 
double occupancy $D=\langle n_{d\uparrow}n_{d\downarrow}\rangle\rightarrow 1-n_d+D$, 
thermopower $S\rightarrow -S$, with $G$ and $K_{\rm e}$ remaining unchanged. The behavior of $S$,
$G$ and $K_{\rm e}$ under ${\rm v}_g\rightarrow -{\rm v}_g$, follows from their definition 
and the behavior of the spectral function, $A(\omega,T)\rightarrow A(-\omega,T))$ 
under ${\rm v}_g\rightarrow -{\rm v}_g$.

Fig.~\ref{figure1} shows the temperature dependence of the dot level occupancy, $n_{d}(T)$, for
gate voltages in the Kondo, mixed valence and empty orbital regimes, for $U/\Gamma = 8$. 
For $U/\Gamma\gg 1$ we use the occupancy at $T=0$ to delineate between the different regimes. 
Specifically, the Kondo regime is defined by gate voltages 
around mid-valley (${\rm v}_g=0$) with $|n_{d}(T=0)-1|\lesssim 0.25$ 
(see caption to Fig.~\ref{figure1}). Similarly, 
the mixed valence and empty orbital regimes are defined by gate voltages corresponding to 
$|n_{d}(T=0)-0.5|\lesssim 0.25$ and $n_{d}(T=0)< 0.25$, respectively (see Fig.~\ref{figure1}). 
In the Kondo regime, a characteristic low temperature scale, the Kondo scale $T_{K}$, can
be defined via \cite{hewson.97}
\begin{equation}
T_{K} = U\sqrt{\frac{\tilde{\Gamma}}{2U}}e^{\frac{\pi \epsilon_{d}(\epsilon_{d}+U)}
{2\tilde{\Gamma} U}} 
= \Gamma\sqrt{\frac{\tilde{u}}{4}}e^{\pi({\rm v}_g^{2}-\tilde{u}^{2}/4)/\tilde{u}},\label{HaldaneTK}
\end{equation}
where $\tilde{u}=U/\Gamma$. The mid-valley Kondo scales for $\tilde{u}=3,6$ and $8$, are 
$T_{K}/\Gamma = 8.2\times 10^{-2}, 1.0\times 10^{-2}$ and $2.64\times 10^{-3}$, respectively. 

Within the FDM approach, 
the thermodynamic value of the dot occupancy $n_{d}={\rm Tr}[\rho n_{d}]$, 
where $\rho$ is the FDM defined in Appendix~\ref{full-density-matrix}, and the value obtained
from the spectral sum rule 
$n_{d}=\sum_{\sigma}\int_{-\infty}^{+\infty} -\frac{1}{\pi}{\rm Im}[F_{\sigma}(\omega,T)]d\omega$, with
$F_{\sigma}$ defined in (\ref{correlationGF}), are identical by construction, as we also verified 
numerically.
Fig.~\ref{figure2} shows the gate voltage dependence of the dot occupancy (and for completeness
also the double occupancy $D=\langle n_{d\uparrow}n_{d\downarrow}\rangle$) 
at a number of temperatures for the strong correlation case $U/\Gamma=8$.

\subsection{Physics of Kondo, mixed valence and empty orbital regimes}
Before presenting the results, a few words are in order concerning the physical significance of the Kondo, mixed valence and
empty orbital regimes for strong Coulomb correlations ($U/\Gamma\gg 1$) on the dot 
(for more detailed information we refer the reader to
Ref.~\onlinecite{hewson.97}). The Kondo regime, $n_{d}\approx 1$, corresponds to the
formation of a localized spin on the dot at intermediate temperatures ($T_{K}\lesssim T\ll \Gamma$).
In this temperature range, physical properties exhibit logarithmic temperature dependences, the
hall-mark of the Kondo effect. At $T<T_{K}$ the localized spin is quenched by the lead electrons,
resulting in a many-body singlet at $T=0$ and a narrow Kondo resonance (of width $T_{K}$)
in the dot spectral density at the Fermi level. Physical properties are characterized by
spin fluctuations on scales $T_{K}\lesssim T\ll\Gamma$, charge fluctuations on a scale 
$T\approx \Gamma$ and renormalized Fermi liquid excitations at $T\ll T_{K}$. 
The dot spectral density is well understood \cite{costi.94}: it has a three peaked structure,
with single-particle charge excitations at $\varepsilon_{d}$ and $\varepsilon_{d}+U$ and a
temperature dependent Kondo resonance at the Fermi level. The mixed valence regime, 
corresponds to gate voltages such that the level $\varepsilon_{d}$ is within $\tilde{\Gamma}$
of the Fermi level. The charge on the dot fluctuates between $n_{d}=0$ and 
$n_{d}=1$ resulting in an average charge $n_{d}\approx 0.5$. The physics is 
governed by quantum mechanical charge fluctuations on a scale
set by $\tilde{\Gamma}$. The empty orbital regime, corresponds to $n_{d}\approx 0$ 
and $\varepsilon_{d}/\Gamma \gg 1$. Physical properties are dominated by charge 
fluctuations, primarily via thermal activation (with an activation energy $\varepsilon_{d}$).
Even though $U/\Gamma \gg 1$, the physics in this regime corresponds to that of a
non-interacting resonant level model with a resonant level of width $\tilde{\Gamma}$ 
at energy $\varepsilon_{d}>0$.

\section{Temperature dependence of transport properties of quantum dots}
\label{Tdependence}
\begin{figure*}[t]
\includegraphics[scale=0.5]{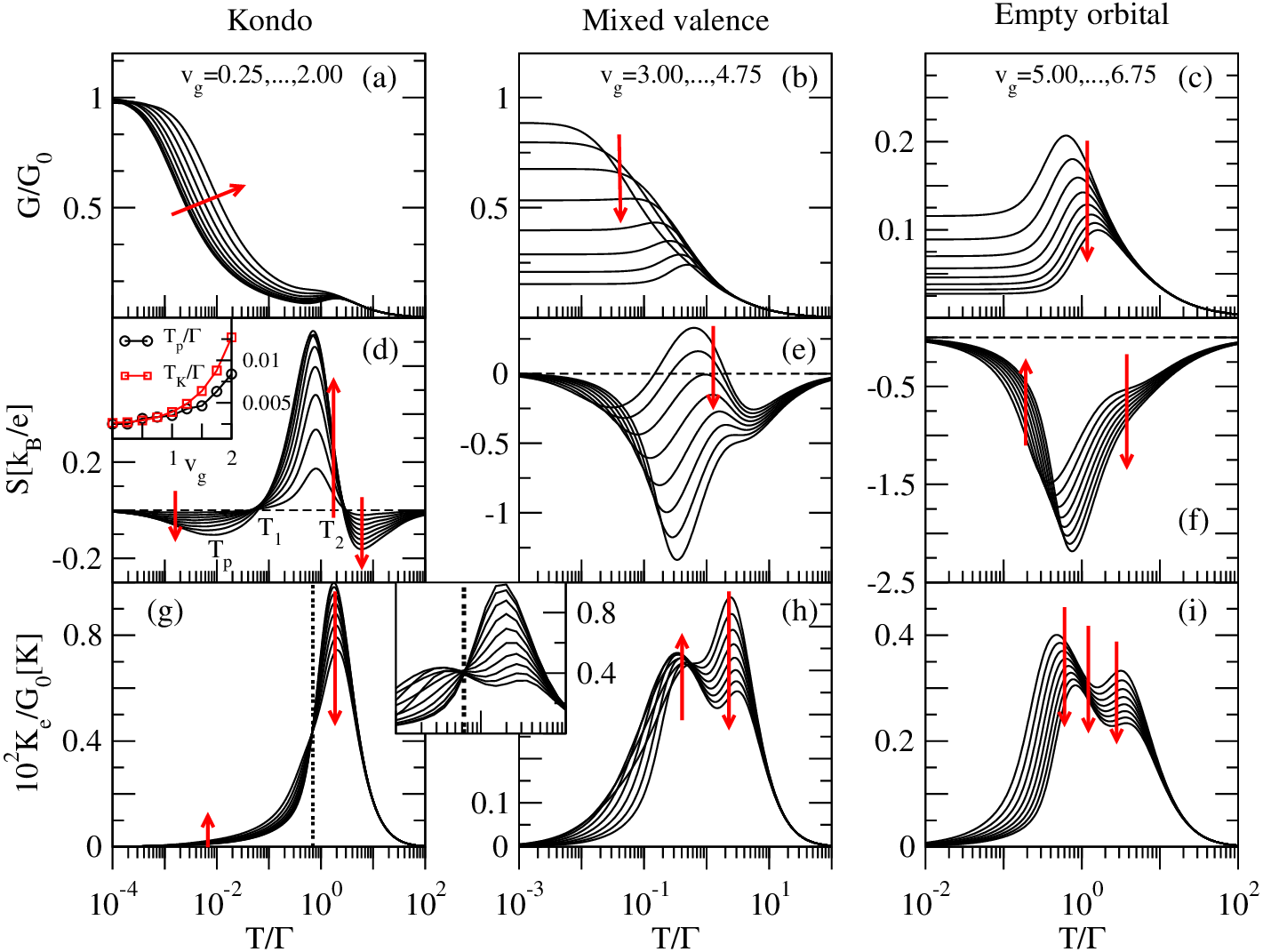}
\caption{  (a-c) The normalized electrical conductance,
$G/G_{0}$, where $G_{0}=2e^{2}/h$, (d-f), the thermopower, $S$, in units of $k_{B}/e$, 
and, (g-i), the normalized electronic contribution to the thermal conductance, 
$K_{\rm e}/G_{0}$, multiplied by a factor $10^2$ for clarity of presentation, 
as function of $T/\Gamma$, in the strongly correlated regime $U/\Gamma=8$ for a range 
of  gate voltages, ${\rm v}_g=(\varepsilon_{d}+U/2)/\Gamma$, in the Kondo 
(first column), mixed valence (second column) and empty orbital 
(third column) regimes. The range of ${\rm v}_g$ is indicated in the top
panels for each regime and the increment used was $0.25$. Arrows indicate 
the evolution of the transport quantities with increasing ${\rm v}_g$.
The inset to (d) compares the Kondo scale $T_{K}$ in the Kondo regime with the
peak position of the low temperature peak, $T_{p}$, in $S$ below the
first sign change at $T_{1}$ as a function of ${\rm v}_g>0$. In the Kondo regime
$T_{K}\approx T_{p}$ and for gate voltages approaching the mixed valence 
regime, the two scales deviate, as expected. 
The inset for $K_{\rm e}$ in (g-h) shows the crossing point
at $T/\Gamma \approx 0.6$ in more detail (vertical dotted line) and the
evolution of the two-peaked structure for gate voltages approaching the
mixed-valence regime.
} \label{figure3}
\end{figure*}

The temperature dependence of transport properties of a quantum dot described by the model 
(\ref{qdot-two-leads}) is shown in Fig.~\ref{figure3} for several values of the gate voltage,
ranging from the Kondo regime (Fig.~\ref{figure3}a,d,g), to the mixed valence (Fig.~\ref{figure3}b,e,h)
and empty orbital (Fig.~\ref{figure3}c,f,i) regimes and for strong
Coulomb correlations on the dot ($U/\Gamma =8$). Moderate to weak correlations are described 
briefly in Sec.~\ref{moderate-correl} and Appendix~\ref{extra-results}.
Depending on the regime, 
the transport properties exhibit different characteristic temperature dependences, which we describe 
in detail below for each transport property in turn.
Here, and in several other figures in the paper,
we use arrows to indicate the evolution, with increasing gate voltage ${\rm v}_{g}>0$ 
about mid-valley (${\rm v}_{g}=0$), of the various transport properties. 
\subsection{Electrical conductance: $G(T)$}
The general trends in the electrical conductance $G(T)$ of Kondo correlated quantum dots are 
well understood\cite{glazman.88,ng.88,costi.94,costi.00,costi.01,izumida.98}: in short,
as $T\rightarrow 0$, the conductance approaches a maximum value (see Fig.~\ref{figure3}a), 
indicating that the quantum 
dot appears ``transparent'' to electrons tunneling through it, and a logarithmic behavior 
around $T_K$ marks the crossover from the weakly coupled regime at $T\gg T_{K}$ to the strongly 
coupled regime at $T\ll T_{K}$. 
An issue, less discussed in the literature, which we point out here, is the
appearance of a finite temperature peak in the conductance, $G(T)/G_{0}$, on
entering the mixed valence regime (see Fig.~\ref{figure3}b). This feature becomes 
particularly pronounced in the empty orbital regime (see Fig.~\ref{figure3}c).
This effect has been observed in experiments on lateral quantum dots
\cite{goldhaber-gordon.98,vanderwiel.00} and a comparison to theoretical calculations 
shows good agreement\cite{koenig.00,costi.03} (see also the discussion of the resistivity
of dilute magnetic impurities in Sec.~\ref{Tdependence-compare}). 
\subsection{Thermopower: $S(T)$}
\label{thermopower-dot-tdep}
\begin{figure}[t]
\includegraphics[scale=0.3]{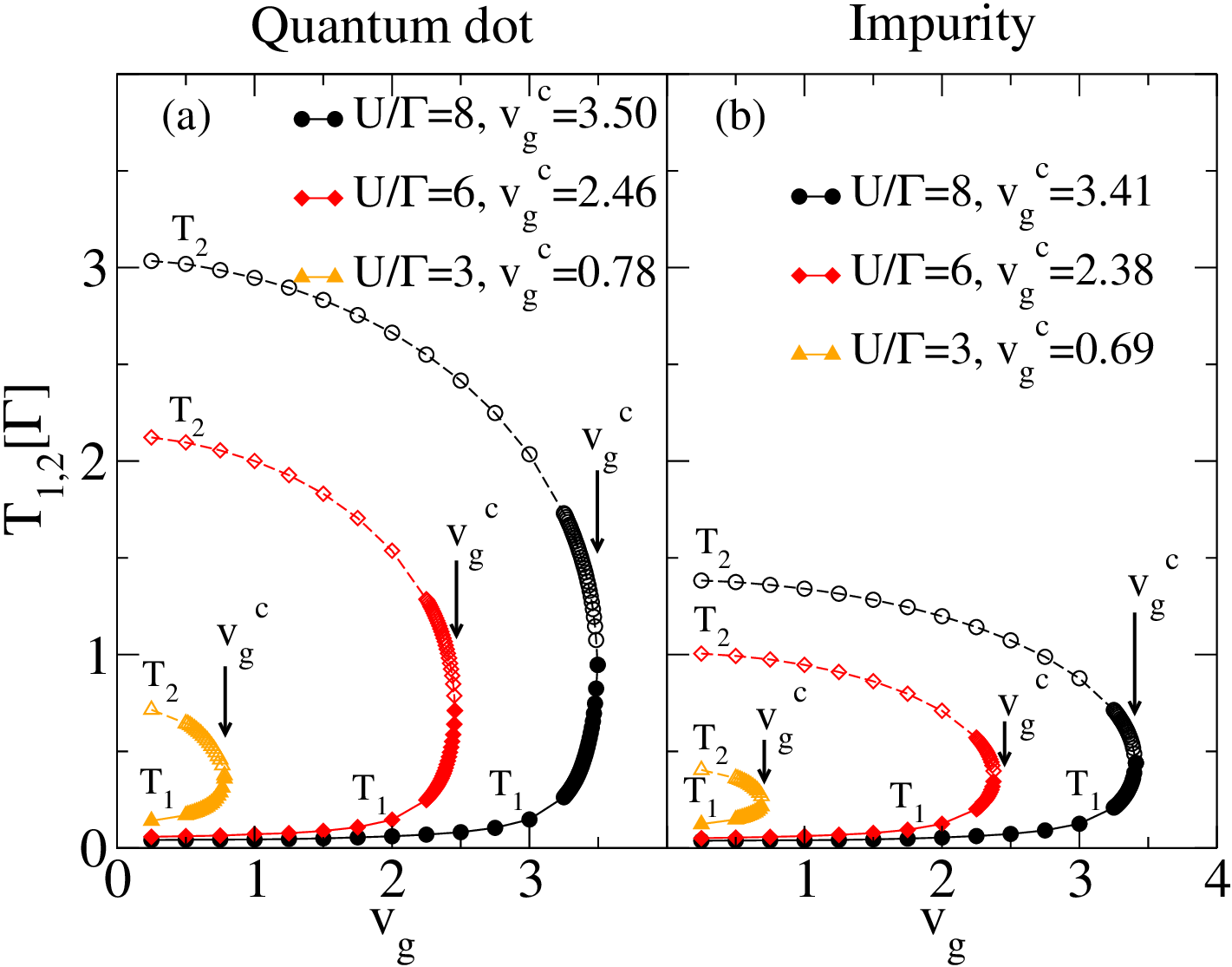}
\caption{  Dependence of the temperatures $T_{1}$ (filled symbols, solid lines) 
and $T_{2}$ (open symbols, dashed lines) at which $S(T)$ changes sign as a 
function of ${\rm v}_g\ge 0.25$ for $U/\Gamma =3, 6,8$ 
(left panel: quantum dot, right panel: magnetic impurity). 
The critical gate voltage, ${\rm v}_g^{c}$, beyond which no sign change occurs in 
$S(T)$ at finite ${\rm v}_g$ is indicated in the legend for each case. 
} 
\label{figure4}
\end{figure}

The thermopower exhibits a particularly interesting temperature
dependence in the Kondo regime, Fig.~\ref{figure3}d, with two sign 
changes at $T=T_{1}({\rm v}_g)$ and $T=T_{2}({\rm v}_g)$, and, correspondingly, 
three extrema at $T=T_{p}$, $T\approx 0.6-0.8\Gamma$ and $T\approx 6\Gamma$. 
The detailed behavior of $T_{1}$ and $T_{2}$ as a function of gate voltage 
will be described below; here, it suffices to note that neither $T_{1}$ 
nor $T_{2}$ are low energy scales, and $T_{2}$ is typically on a scale 
of order $\Gamma$ (see Fig.~\ref{figure3}d and Fig.~\ref{figure4}a below). 
The low temperature ``Kondo'' peak in $S(T)$ at $T=T_{p}$ is
found to scale with $T_{K}$ (as defined in Eq.~(\ref{HaldaneTK})),
as shown in the inset to  
Fig.~\ref{figure3}d. Thus, in contrast to $T_{1}$ and $T_{2}$, $T_{p}$ can 
be considered a low energy scale in the Kondo regime. 
The central positive peak in $S(T)$ 
first grows with positive magnitude on moving away from the Kondo regime (Fig.~\ref{figure3}d)
and then decreases in magnitude in Fig.~\ref{figure3}e on entering the mixed valence 
regime. Simultaneously, the ``Kondo'' peak in $S(T)$ acquires a large negative value 
while merging with the high energy  (negative) peak at $T\approx 6\Gamma$ 
on entering the mixed valence regime (Fig.~\ref{figure3}e). Well into the mixed valence
regime, the thermopower exhibits a single negative peak on a scale $\Gamma$ with a distinct 
shoulder at higher temperatures due to the peak at $T\approx 6\Gamma$. 
This picture continues 
to hold in the empty orbital regime (see Fig.~\ref{figure3}f), with the shoulder 
at $T\approx 6\Gamma$ having almost disappeared. The thermopower remains negative for all gate
voltages ${\rm v}_g>0$ in this regime. 

The above behavior in the temperature dependence of the thermopower 
in the Kondo regime is explained in terms of the structure of the 
single-particle excitations in $A(\omega,T)$. At low temperatures, 
a Sommerfeld expansion for 
$S(T)$ gives\cite{costi.94}
\begin{equation}
S(T) = -\frac{k_{B}}{|e|}\frac{\pi^{2}}{3}k_{B}T\frac{1}{A(0,T)}\,
\frac{\partial A}{\partial\omega}|_{\omega=0},
\label{sommerfeld-exp}
\end{equation}
showing that the sign of the thermopower depends on the slope of the spectral
density at the Fermi level. For $T\ll T_{K}$ and ${\rm v}_g>0$, the Kondo resonance
lies above the Fermi level, so the slope of the spectral density at the Fermi level
is positive, resulting in a negative thermopower. This remains true on further increasing the
temperature, but as shown in Ref.~\onlinecite{costi.94}, eventually the Kondo resonance
is suppressed at $T>T_{K}$ resulting in a negative slope of the spectral density at
$\omega=0$ for ${\rm v}_g>0$ (with the opposite being true for ${\rm v}_g<0$). Consequently,
the thermopower changes sign at the temperature $T_{1}$ which roughly corresponds to
the temperature at which the Kondo resonance vanishes. At $T\gg T_{K}$, the 
determining factor for the sign of the thermopower is no longer the slope of
the spectral function at 
$\omega=0$, but the number of states available below or above the Fermi level.
These determine the overall sign of the transport integral $I_{1}$ in the expression for
the thermopower in Eq.~(\ref{transport-expressions-S}). For ${\rm v}_g>0$, there are $n_{d}/2 < 0.5$ 
states below the Fermi level and $1-n_{d}/2 > 0.5$ states above the Fermi level. 
Consequently, the integral of $-(\partial f/\partial \omega)\omega A(\omega,T)$ 
for $\omega >0$ is greater than its counterpart for $\omega<0$, so $I_{1}>0$ 
and the thermopower is again  negative at $T\gg T_{K}$. This occurs at $T=T_{2}$, 
which is found to be of order $\Gamma$ (see below). Due to the factor $1/T$ coming from the derivative of the
Fermi function in $I_{1}$, the negative thermopower at $T>T_{2}$ acquires a maximum
negative value and then decreases as $1/T$ at $T\gg \Gamma$, exhibiting no further sign 
changes, as confirmed also numerically. We note that, away from half-filling (${\rm v}_g=0$),  
the modified second order perturbation in $U$ 
approach\cite{craco.99,martin-rodero.82} gives an incorrect sign for the 
slope of the spectral density at the Fermi level in the Kondo regime.
This results in a wrong sign for the thermopower at $T<T_{K}$  in the
Kondo regime\cite{dong.02} compared to our NRG calculations 
(which agree with those of Ref.~\onlinecite{yoshida.09}). 
Approximate approaches using an infinite $U$ Anderson model 
\cite{boese.01,franco.08} could also not access the low temperature Kondo regime. 

The sign changes of the thermopower of strongly correlated 
quantum dots at the temperatures $T_{1}$ and $T_{2}$ in the Kondo regime, are
particularly interesting. They provide
a ``smoking gun'' signature for Kondo behavior in quantum dots, and could be used in future
experiments as sensitive probes of strong correlations and Kondo physics. 
It is therefore interesting to give a detailed characterization of the dependence of
$T_{1}$ and $T_{2}$ on gate voltage and interaction strength $U/\Gamma$. We show in
Fig.~\ref{figure4}a-b the loci of $T_{1}$ and $T_{2}$ as a function of ${\rm v}_g>0$ for 
quantum dots and magnetic impurities for three interaction strengths. Although
$S(T)$ vanishes at ${\rm v}_g=0$, $T_{1}$ and $T_{2}$ have finite limiting
values there. These are difficult to determine numerically due to the vanishingly
small thermopower in this limit, and they are difficult to obtain analytically, since
$T_{1}$ and $T_{2}$ lie outside the Fermi liquid regime where analytic calculations
are possible. Estimates of these values at the smallest gate voltage are close to the 
limiting values. They are tabulated in Table~\ref{tableTK}, together with the relevant 
Kondo scales at mid-valley. Whereas the limiting values of $T_{1}$ are 
comparable for both quantum dots and magnetic
impurities, the limiting values of $T_{2}$ for quantum dots are approximately twice larger
than for magnetic impurities. By carrying out additional calculations, using a finer grid
of gate voltages, we determined the critical gate voltages ${\rm v}_g^{c}(U/\Gamma)$,
beyond which no sign change occurs (indicated in Fig.~\ref{figure4}a-b). For each value
of $U/\Gamma$, we find that ${\rm v}_g^{c}$ corresponds to entering the mixed valence regime, i.e.
${\rm v}_g^{c}$ corresponds to a local level position $\varepsilon_{d}\approx -\Gamma/2$ 
in the single channel Anderson model.
\vspace{0.5cm}
\begin{table}[t]
\begin{center}
\begin{tabular}{|c|c|c|c|}
   \hline
     $U/\Gamma$ & $T_{K}({\rm v}_g=0)/\Gamma$ & 
   $T_{1}({\rm v}_g=0.25)/\Gamma$ & $T_{2}({\rm v}_g=0.25)/\Gamma$\\
   \hline
     $8$ & $2.64 \times 10^{-3}$ & $0.044$ ($0.04$) & $3.04$ ($1.39$)\\
   \hline
     $6$ & $1.10\times 10^{-2}$  & $0.056$ ($0.05$) & $2.12$ ($1.00$)\\
   \hline
     $3$ & $8.20\times 10^{-2}$  & $0.13$  ($0.12$) & $0.71$ ($0.41$)\\
   \hline
\end{tabular}
\end{center}
\caption{  Kondo temperature $T_{K}$ (in units of $\Gamma$) at mid-valley 
${\rm v}_g=0$ (symmetric point) for $U/\Gamma=8,6,3$.  Also shown are the 
temperatures $T_{1}$ and $T_{2}$ (in units of $\Gamma$) at which the 
thermopower, $S(T)$, changes sign at ${\rm v}_g=0.25$ (the smallest ${\rm v}_g$ 
studied). The numbers in brackets are the corresponding temperatures 
for magnetic impurities.
}
\label{tableTK}
\end{table}
\subsection{Thermal conductance: $K_{\rm e}(T)$}
The electronic contribution to the thermal 
conductance of a strongly correlated quantum dot, shown in Fig.~\ref{figure3}g-i, also 
exhibits interesting behavior: a crossing point at $T\approx \Gamma$ is found 
in the Kondo regime and for gate voltages approaching the mixed valence regime 
(Fig.~\ref{figure3}g and inset). Such (approximate) crossing points are
typical signatures of strong correlations and are well known in other 
contexts, including $^{3}{\rm He}$ and heavy fermions\cite{vollhardt.97},
dissipative two-level systems \cite{costi.99} and doped Mott-insulators\cite{laad.01}. 
On entering the mixed valence and empty orbital regimes (Fig.~\ref{figure3}h-i), 
two-peaks develop on either side of the crossing point (the lower peak being at
$T\lesssim 0.5\Gamma$ and the upper one at $T\gtrsim 2\Gamma$). These qualitative features
in $K_{\rm e}(T)$ can be related to $A(\omega,T)$, as in the case of $S(T)$ 
(see also Sec.~\ref{compare-thermal-cond}).
\subsection{Moderate to weak correlations}
\label{moderate-correl}
The effect of reducing correlations to a moderate value, $U/\Gamma = 3$,
is shown in Fig.~\ref{figure11} of Appendix~\ref{extra-results}: the
trends are similar to those described above, with a significantly diminished Kondo 
regime. In particular, the evolution with gate voltage 
of $G(T)$ is similar to that in the strongly
correlated case (see Fig.~\ref{figure11}a-c) 
and the thermopower exhibits two sign changes as a function of temperature
in the Kondo regime (Fig.~\ref{figure11}d), with a rapid evolution to a single negative peak
in the mixed valence and empty orbital regimes (Fig.~\ref{figure11}e-f).
However, the crossing point in $K_{\rm e}$ in the Kondo regime 
becomes less evident for moderate correlations (Fig.~\ref{figure11}g), 
and, the two-peaked structure for $K_{\rm e}$ in the mixed valence and empty orbital 
regimes is replaced by a single peak with a shoulder (Fig.~\ref{figure11}h-i).

These general trends, for correlated quantum dots, contrast with those for weakly correlated 
quantum dots, shown in Fig.~\ref{figure12}a-c of Appendix~\ref{extra-results} for $U/\Gamma=1$. 
These exhibit no sign change in the thermopower for any gate voltage ${\rm v}_g>0$. Similarly, the thermal 
conductance for weakly correlated quantum dots shows no crossing point, 
exhibiting only a single finite temperature peak.
\subsection{High temperature asymptotics}
\label{asymptotics}
The FDM approach allows us to easily 
investigate the high temperature asymptotics of transport properties. As we discuss 
also in the context of dilute magnetic impurities in 
Sec.~\ref{Tdependence-compare} below, earlier transport
calculations \cite{costi.94} could not discern the highest temperature peak in 
$S(T)$ (occurring at $T\approx 6\Gamma$ for $U/\Gamma=8$, see Fig.~\ref{figure3}d), nor
the peak in the thermal conductivity (see discussion in Sec.~\ref{Tdependence-compare} below). 
Here, we are able to do so. In addition, the numerical calculations recover
the high temperature asymptotics of the transport properties: $G(T)\sim 1/T$, $S(T)\sim 1/T$ and 
$K_{\rm e}(T)\sim 1/T^{2}$ for $T\gg \Gamma$. Note that, for the Anderson model, the logarithmic
corrections in the Kondo regime occur at intermediate temperatures $T_{K}<<T<<\Gamma$: the
corrections at $T\gg \Gamma$ go over to the above power laws.
\section{Comparison with dilute magnetic impurities}
\label{Tdependence-compare}
It is interesting to quantify the differences in the transport properties of quantum dots
given by (\ref{transport-expressions-G}-\ref{transport-expressions-K}) 
with the analogous transport properties of dilute magnetic impurities given by 
(\ref{imp1}-\ref{imp3}). This is shown in Fig.~\ref{figure5} for the temperature 
dependence of transport properties in the Kondo, mixed valence and empty 
orbital regimes for $U/\Gamma=6$. 

\subsection{Comparison of $G(T)$ and $\rho(T)$}
In the Kondo regime, and for temperatures $T<<\Gamma$, the conductance of a quantum dot
is a universal function of $T/T_{K}$, i.e. $G(T)/G(0)=f(T/T_{K})$ (e.g. see Ref.~\onlinecite{costi.00}). 
The same holds for the
analogous quantity for dilute magnetic impurities, namely the resistivity, i.e. 
$\rho(T)/\rho(0)=f'(T/T_{K})$ (e.g. see Ref.~\onlinecite{costi.94}). 
Since $G$ and $\rho$ are different physical quantities,
the functions $f$ and $f'$ are different and they cannot be made to 
coincide by using a common Kondo scale $T_{K}$ (e.g. the
Kondo scale defined in Eq.~\ref{HaldaneTK}). This is seen in Fig.~\ref{figure5}a, which shows that the
conductance curves for quantum dots are shifted in temperature, on a logarithmic scale, 
relative to the resistivity curves of magnetic impurities. The two functions $f$ and $f'$ 
are rigorously identical only in the Fermi liquid regime $T\ll T_{K}$. Experimentally, however, the
accessible range of temperatures is that around $T\approx T_{K}$, say one decade below and one
decade above $T_{K}$. For this region of temperatures, the two functions $f$ and $f'$ can be made to coincide 
by redefining them as new functions $\tilde{f}$ and $\tilde{f}'$, respectively,
with {\em different} respective Kondo scales, $T^{\rm G}_{K}$ and $T^{\rho}_{K}$ 
such that $\tilde{f}(T/T^{\rm G}_{K}=1)=\tilde{f}'(T/T^{\rho}_{K}=1)=1/2$, see 
Ref.~\onlinecite{costi.00}. 
In the mixed valence and empty orbital regimes, Fig.~\ref{figure5}b-c shows 
that the conductance of a quantum dot differs significantly from the resistivity of 
magnetic impurities (with significant deviations at $T\gtrsim 0.1\Gamma$). In particular, the aforementioned 
finite temperature peak in the conductance of a quantum dot is absent in the resistivity
of magnetic impurities. A signature of this peak in $\rho(T)$ is seen at most in 
the Kondo regime at temperatures of order $\Gamma$ (see Fig.~\ref{figure5}a) and is absent
in the mixed valence and empty orbital regimes. These differences to the quantum dot case, arise,
as described in Sec.~\ref{transport}, due to the different way in which the spectral function appears 
in the respective transport integrals. These differences reflect also the absence of universality 
outside the Kondo regime.

\subsection{Comparison of thermopowers: $S(T)$}
\label{thermopower-compare}
\begin{figure*}[t]
\includegraphics[scale=0.5]{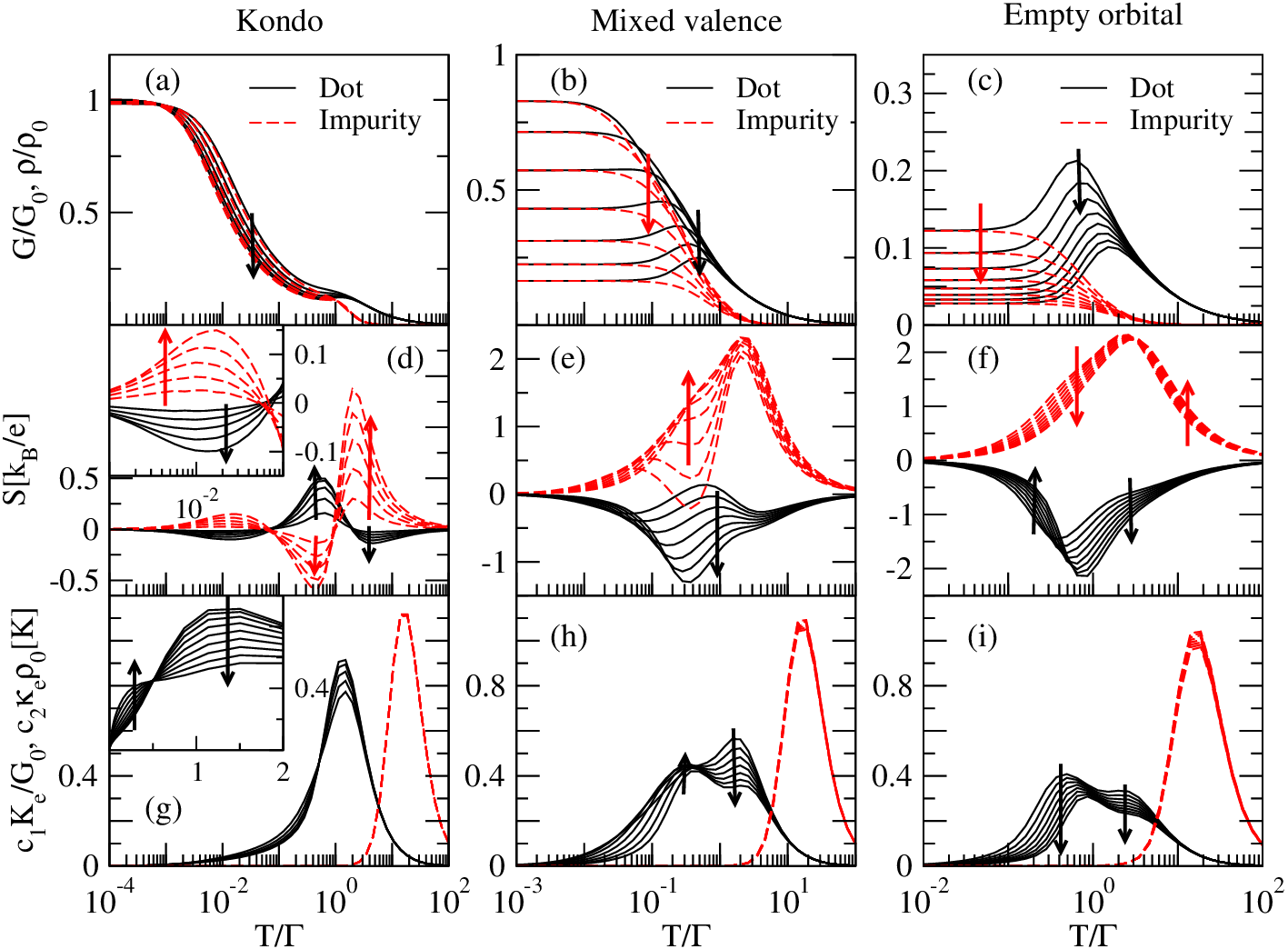}
\caption{Comparison of transport properties for quantum dots (solid lines) and magnetic 
impurities (dashed lines) in the strongly correlated regime $U/\Gamma=6$. 
(a-c): normalized electrical conductance $G(T)/G_{0}$ (quantum dot) and 
normalized resistivities $\rho/\rho_{0}$ (impurity), where $\rho_{0}$ is defined 
in (\ref{unitary-rho}). (d-f): thermopower $S(T)$ for quantum dots and impurities
(inset to (d) shows the low temperature Kondo peak in the thermopower in more detail). 
(g-i): electronic contribution to the thermal conductance $c_{1}K_{\rm e}/G_{0}$ (quantum dot) 
and thermal conductivity $c_{2}\kappa_{\rm e}\rho_{0}$ (impurity), rescaled
by $G_{0}$ and $\rho_{0}$ respectively, so that the same unit ($K$) applies to both
cases. The numerical factors $c_{1}=10^{2}$ and $c_{2}=5\times 10^{-4}$ are included 
for clarity of presentation. The inset in (g) for 
$K_{\rm e}(T)/G_{0}$ shows the crossing point in $K_{\rm e}(T)/G_{0}$ at 
$T/\Gamma \approx 0.6$ in more detail and the evolution of the second peak in the
thermal conductance of the quantum dot as the mixed valence regime is approached
(with inclusion of 4 additional gate voltages ${\rm v}_g=1.50,\dots,2.25$ 
in the Kondo regime). The range of gate voltages is ${\rm v}_{g}=0.25,\dots,1.25$ (Kondo),
${\rm v}_{g}=2.25,\dots,3.75$ (mixed valence), and ${\rm v}_{g}=4.0,\dots,5.75$ (empty orbital).
} \label{figure5}
\end{figure*}

In Fig.~\ref{figure5}d-f we see that, up to an overall sign change, due to
$A(\omega,T)$ appearing differently in the transport integrals as explained in Sec.~\ref{transport},
the thermopower of magnetic impurities behaves in a qualitatively similar way to that 
of a quantum dot, with two sign changes at $T_{1}({\rm v}_g)$ and $T_{2}({v}_g)$ (shown
in Fig.~\ref{figure4}b) and three extrema. In the Kondo regime, the position, $T_{p}$, 
of the Kondo enhanced peak in the thermopower of magnetic impurities is found to scale 
with $T_{K}$, just as for the quantum dot case (see Sec.~\ref{thermopower-dot-tdep}).
A significant difference between $S(T)$ for magnetic impurities and quantum dots is 
the much larger high temperature peak (at $T>T_{2}$) for the former in the Kondo regime 
(by as much as a factor $5$, see Fig.~\ref{figure5}d). This difference holds to some 
extent also in the mixed valence regime (Fig.~\ref{figure5}e). In the empty orbital
regimes the thermopowers show a single peak at $T\approx \Gamma$ with a similar magnitude
for both cases (Fig.~\ref{figure5}f). 

\subsection{Comparison of $K_{\rm e}(T)$ and $\kappa_{\rm e}(T)$}
\label{compare-thermal-cond}
The electronic contribution to the thermal conductivity of magnetic impurities, $\kappa_{\rm e}$, 
shows significant differences to the corresponding thermal conductance of
quantum dots, $K_{\rm e}$, see Fig.~\ref{figure5}g-i. For example, whereas $K_{\rm e}$ exhibits interesting
structure with either one (in the Kondo regime) or two (in the mixed valence and empty orbital regimes)
peaks around $T\approx \Gamma$, $\kappa_{\rm e}$ only exhibits a single peak in all regimes, and this
peak occurs at a much larger temperature $T\gg \Gamma$. The reason for the latter difference 
is the following: the main contribution to
the thermal conductance and thermal conductivity come from the integrals $I_{2}$ and $M_{2}$ in
Eq.~(\ref{trans-dot}) and Eq.~(\ref{trans-imp}),respectively, which involve integrals
of $-(\partial f/\partial \omega)\omega^{2}A(\omega,T)$ and 
$-(\partial f/\partial \omega)\omega^{2}/A(\omega,T)$, respectively. 
For the former, the peaks in $A(\omega,T)$ at 
$\varepsilon_{d}$ and $\varepsilon_{d}+U$ result in a peak in the 
integrand at $|\omega|\gtrsim U/2\gtrsim \Gamma$,
whereas in the latter, the dips in $1/A(\omega,T)$ at $\varepsilon_{d}$ and $\varepsilon_{d}+U$ 
shift the main contribution to the integral to much higher energies $|\omega|\gg U/2 $. Correspondingly,
the temperature of the peaks in $K_{\rm e}$ in the former are at $T\lesssim U/2$ and for the latter are
at $T\gg U/2>\Gamma$, in agreement with the numerical results. The existence of two peaks in
$K_{e}$ in the mixed valence and empty orbital regimes as opposed to a single peak in the Kondo regime is
also easily explained: the two peaks reflect the sampling of the two incoherent features
at $\varepsilon_{d}$ and $\varepsilon_{d}+U$ in $\omega^{2}A(\omega,T)$ appearing in the moment $I_{2}$ for
$K_{\rm e}$. In the Kondo regime, these excitations lie close to each other and only one peak results.
Similarly, the single peak in $\kappa_{\rm e}$ for all regimes results from the strong suppression
of the above incoherent excitations in $1/A(\omega,T)$ appearing in the moment $M_{2}$.

\section{Figure of merit, power factor and Lorenz number}
\label{zt}
A measure of the thermoelectric efficiency of a quantum dot device is the dimensionless
figure of merit defined by $ZT=GS^{2}T/(K_{\rm e}+K_{\rm ph})$, where $K_{\rm ph}$ is the 
phonon contribution to the thermal conductance. Hence for high efficiency, one requires
either large $S$ or small total thermal conductance relative to electrical conductance or
both conditions simultaneously. A calculation of $ZT$ for quantum dot systems would 
therefore require knowledge of the material specific phonon contribution to the thermal 
conductance $K_{\rm ph}$. Similarly, for magnetic impurity
systems a calculation of the dimensionless figure of merit, 
$ZT=\sigma S^{2}T/(\kappa_{\rm e}+\kappa_{\rm ph})$,
would require knowledge of the material specific phonon contribution to the thermal 
conductivity $\kappa_{\rm ph}$. This is outside the scope of the present paper, 
so instead we show in Fig.~\ref{figure6}a-c results at $U/\Gamma=8$ for the 
quantity $ZT_{0}=GS^{2}T/K_{\rm e}$, 
for quantum dots, and $ZT_{0}=\sigma S^{2}T/\kappa_{\rm e}$, for magnetic impurities (with 
the latter being depicted on the negative axis for clarity). In addition, we also show in Fig.~\ref{figure6}d-f
an appropriate rescaled power factor ($PF_{0}=S^{2}G/G_{0}$ for quantum dots, 
and $PF_{0}=S^{2}\sigma/\sigma_{0}$ for magnetic impurity systems). This is another useful
measure of an efficient thermoelectric system, by-passing lack of knowledge of the 
total thermal conductances (conductivities). 

\subsection{Figure of merit}
Since, in the low temperature limit, the thermopower of the Anderson model 
vanishes linearly with temperature in all regimes, a significant figure of merit is found only 
at finite temperature, as seen in Fig.~\ref{figure6}a-c.

In the Kondo regime, Fig.~\ref{figure6}a, enhanced regions of
$ZT_{0}$ are found in four temperature regions, (i), at $T\approx T_{p}$, due
the low temperature Kondo enhancement of the thermopower, however the magnitude of $ZT_{0}$
is tiny (see inset to  Fig.~\ref{figure6}a), (ii), at temperatures of order $\Gamma$ 
in the region $T_{1}<T<T_{2}$, where $ZT_{0}$ can be of order $0.2-0.3$ for both quantum dots and
magnetic impurities, (iii), at temperatures $T>T_{2}$, with enhancements comparable to those for region (ii),
and, (iv), in the asymptotic region $T\gg \Gamma$, where $ZT_{0}$ saturates to a finite
value which is larger for quantum dots than for magnetic impurities (discussed below).

The behavior of the figure of merit in the mixed valence and empty orbital regimes is 
complicated, see Fig.~\ref{figure6}b-c. In the mixed valence regime, significant enhancements are found, 
for quantum dots, at temperatures somewhat below $\Gamma$, see Fig.~\ref{figure6}b, and in the 
asymptotic regime $T\gg \Gamma$. Similar enhancements are found also for the empty orbital case 
(Fig.~\ref{figure6}c). For magnetic impurities, similar enhancements to quantum dots are
found on temperature scale of order $\Gamma$, but at $T\gg\Gamma$ the enhancements are much
smaller than for quantum dots (see Fig.~\ref{figure6}b-c). The latter effect is due to
the much larger thermal conductivities (even at higher temperatures) of magnetic impurities as 
compared to those of quantum dots (see discussion above and Fig.~\ref{figure6}g-i).

\subsection{Power factor}
The power factor $PF_{0}$ is enhanced in the same regimes (i)-(iii) as the figure of merit,
see Fig.~\ref{figure6}e-f, but vanishes as $1/T^{3}$ in the limit $T\gg \Gamma$ (using
the asymptotic behavior of $S$ and $G$ from Sec.~{\ref{asymptotics}}). In the Kondo
regime, $PF_{0}$ exhibits a much larger peak above $T_{2}$ for magnetic impurities 
as compared to quantum dots (see Fig.~\ref{figure6}d). This reflects the observation made above 
(Sec~\ref{thermopower-compare}) that the highest temperature peak in $S(T)$ at $T>T_{2}$ 
for magnetic impurities is significantly enhanced as compared to that of quantum dots. For 
quantum dots, the main enhancement in $PF_{0}$ in the Kondo regime is in the range $T<T_{2}$.
These trends differ little from those observed in the mixed valence and empty orbital regimes 
for both quantum dots and magnetic impurities (see Fig.~\ref{figure6}e-f).

\subsection{Wiedemann-Franz law and Lorenz number}
\label{wf-lorenz} 
\begin{figure*}[t]
\includegraphics[scale=0.5]{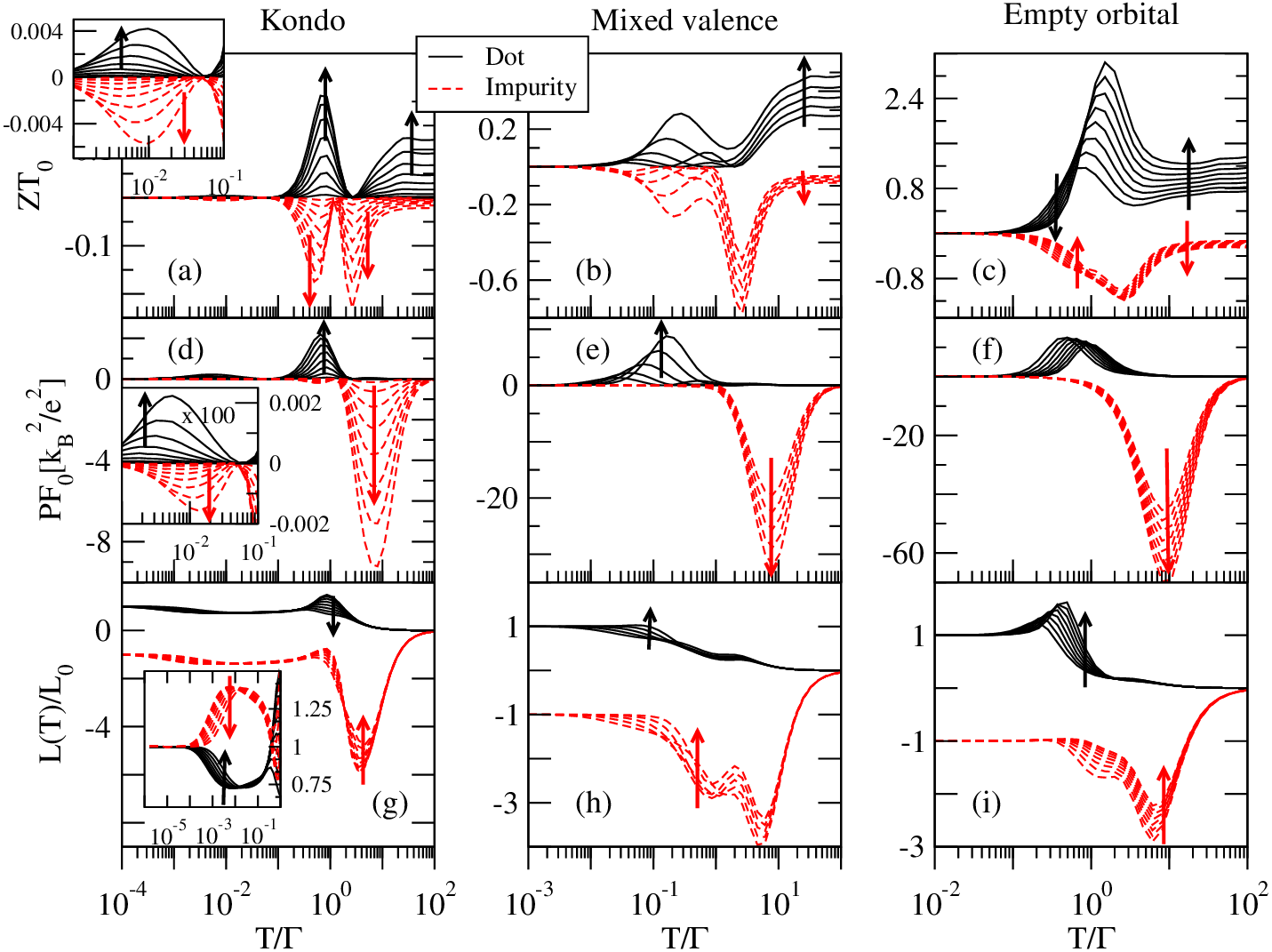}
\caption{Temperature dependence of, (a-c), the ``figure of merit'', $ZT_{0}$, 
(d-f), the ``power factor'', $PF_{0}$, and, (g-i), the Lorenz number ratio 
$L(T)/L_{0}$ for a quantum dot (solid lines, positive y-axis) 
and for magnetic impurities (dashed lines, shown on the negative y-axis for clarity). 
Results are for $U/\Gamma=8$ and the same range of dimensionless gate voltages as in Fig.~\ref{figure3}. 
For quantum dots we define $ZT_{0}=GS^{2}T/K_{\rm e}$,
$PF_{0}=S^{2}G/G_{0}$ and $L(T)/L_{0}=K_{\rm e}/G T$ with $L_{0}=\pi^{2}k_{B}^{2}/3e^{2}$. 
The corresponding quantities for magnetic impurities are defined by 
$ZT_{0}=\sigma S^{2}T/\kappa_{\rm e}$, $PF_{0}=S^{2}\sigma/\sigma_{0}$ and 
$L(T)/L_{0}=(\kappa_{\rm e}/\sigma T)$ where $\sigma=1/\rho$ and $\sigma_{0}=1/\rho_{0}$  . 
Arrows indicate the evolution of the transport quantities with 
increasing ${\rm v}_g>0$. Insets in (a) and (d) for the Kondo regime, 
show the low temperature peak in
$ZT_{0}$ and $PF_{0}$ in the vicinity of $T_{K}({\rm v}_{g})\gtrsim 2.6\times 10^{-3}\Gamma$ 
(Table~\ref{tableTK}). The inset in (g) for the Lorenz number shows the 
deviations from the Wiedemann-Franz law in the region around $T_{K}({\rm v}_{g})$. 
In this inset, $L(T)/L_{0}$ is shown on the positive y-axis 
for both impurity and quantum dot cases.
}\label{figure6}
\end{figure*}

We comment on the enhancement of $ZT_{0}$ in the region $T\gg \Gamma$, which can result in 
$ZT_{0}>1$ (e.g. in the mixed valence and empty orbital cases). This enhancement reflects a
violation of the Wiedemann-Franz law at $T\gg \Gamma$. The latter states that the thermal
conductance (conductivity) is proportional to the electrical conductance (conductivity) multiplied
by temperature, i.e. that the Lorenz number $L(T)$, defined for quantum dots by
\begin{equation} 
L(T) = K_{\rm e}(T)/TG(T),\label{wf-qdots}\\
\end{equation}
and for magnetic impurities by
\begin{equation} 
L(T)  = \kappa_{\rm e}(T)/T\sigma(T)\label{wf-impurity},
\end{equation}
is independent of temperature and takes on the universal value $L_{0}={\pi^2 k_{B}^{2}}/{3e^{2}}$.
Since, $ZT_{0}=S^{2}/L(T)$, a significant reduction of $L(T)/L_{0}$ can 
result in an enhancement of $ZT_{0}$. In Fig.~\ref{figure6}g-i we see that $L(T)/L_{0}$ is much 
suppressed at $T\gg \Gamma$, thereby allowing for significant enhancements in $ZT_{0}$ in this
limit. This enhancement is seen for all regimes, especially for the mixed valence and empty orbital
regimes. We note, however, that from Fig.~\ref{figure6}g-i (and inset), 
the Wiedemann-Franz law is, on the whole, reasonably well satisfied at temperatures 
$T\ll \Gamma$, and becomes exact in the Fermi liquid regime \cite{costi.94}(for 
other violations of the Wiedemann-Franz law see Ref~\onlinecite{kubala.08}).
\section{Universal scaling functions for thermal transport through quantum dots}
\label{scaling}
\begin{figure}[t]
\includegraphics[scale=0.35]{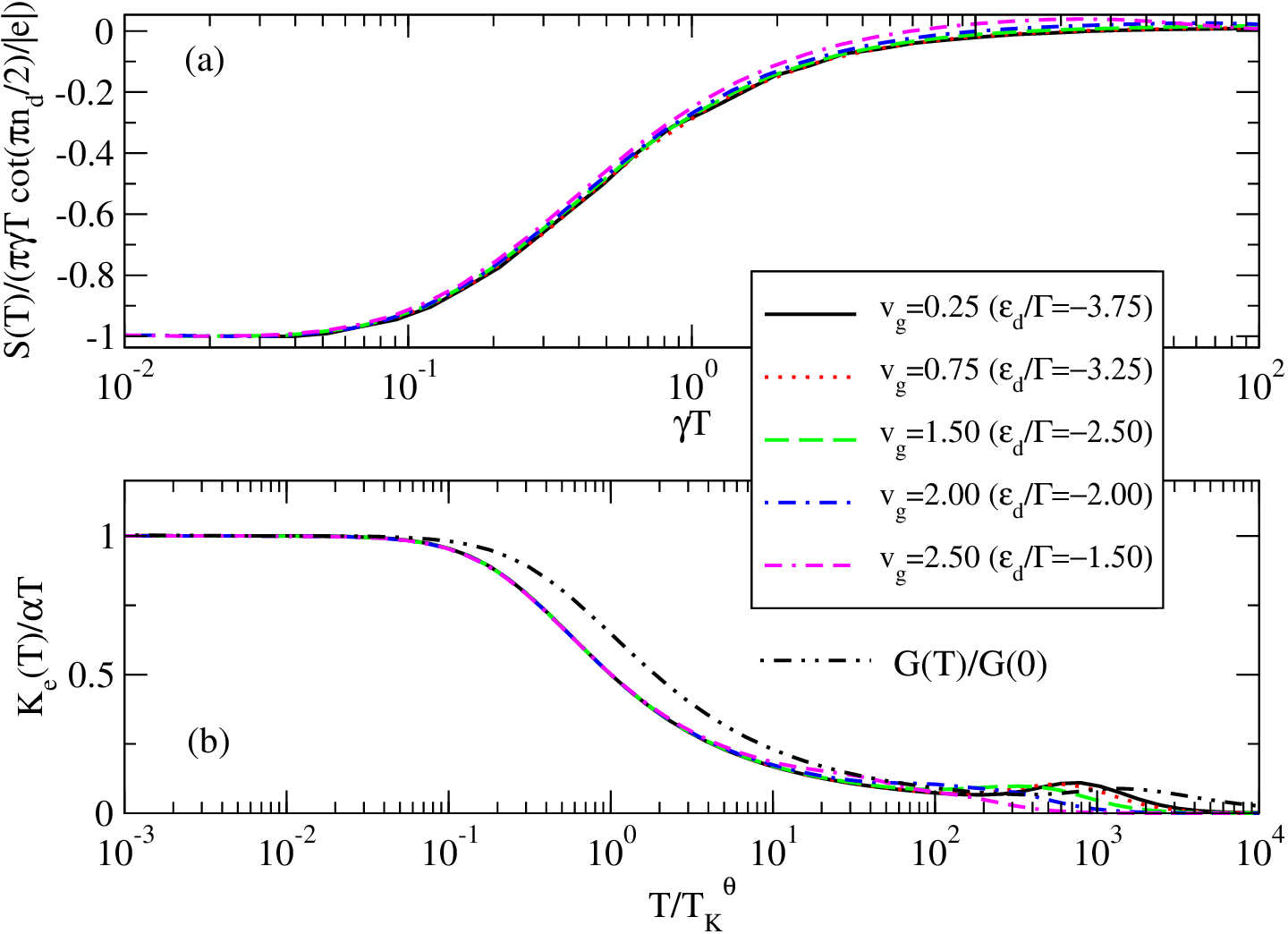}
\caption{  (a) $S(T)/T$ for a quantum dot, scaled by its limiting low temperature 
absolute value $\pi\gamma\cot(\pi n_{d}/2)/|e|$ versus $\gamma T\sim T/T_{K}$ for 
$U/\Gamma = 8$ and a range of gate voltages in the Kondo regime. Here, $\gamma\sim 1/T_{K}$
\cite{specific-heat-note}. (b) the thermal conductance $K_{\rm e}(T)$, scaled by $\alpha T$, 
versus $T/T_{K}^{\theta}$, for parameters as in (a), with
$\alpha$ defined in (\ref{alpha}) and $T_{K}^{\theta}$ a Kondo scale defined in (\ref{tk-theta}).
The electrical conductance $G(T)/G(0)$ versus $T/T_{K}^{\theta}$ at ${\rm v}_g=0.25$ 
is also shown. The Kondo scale defined by $G(T=T_{K})=G(0)/2$ is related to $T_{K}^{\theta}$
by $T_{K}\approx 1.9T_{K}^{\theta}$. 
} \label{figure7}
\end{figure}
By analogy to the scaling properties of the electrical conductance $G(T)/G(0)=f(T/T_{K})$, where
$f$ is a universal function of $t=T/T_{K}$ in the Kondo regime, it is interesting to establish 
to what extent such scaling is present in the thermopower, $S(T)$, and the thermal 
conductance, $K_{\rm e}(T)$, of strongly correlated quantum dots. We investigate this here, 
for $U/\Gamma=8$ and for values of the gate voltage in the Kondo regime (see Fig.~\ref{figure7}).

In the Fermi liquid regime, $T\ll T_{K}$, we have \cite{costi.94} 
\begin{equation}
S(T) = -\frac{\pi\gamma T}{|e|}\cot(\pi n_{d}/2).
\label{thermopower-fl}
\end{equation}
Scaling can therefore be expected for $S(T)/T$, once the occupancy (and gate voltage) dependent 
factor $\cot(\pi n_{d}/2)$ is scaled out. In the above, $\gamma\sim 1/T_{K}$ is a measure
of the inverse Kondo scale and can be extracted\cite{specific-heat-note} from the numerical value of 
$\lim_{T\rightarrow 0}|S(T)|/T$ using Eq.~(\ref{thermopower-fl})
and the calculated values of $n_{d}$ from Fig.~\ref{figure1}. 
We see from Fig.~\ref{figure7}a that $s(T/T_{K})=|e|S(T)/\pi\gamma T\cot(\pi n_{d}/2)$
does indeed scale with $\gamma T \sim T/T_{K}$ for a range of gate voltages in the 
Kondo regime. This scaling extends up to temperatures comparable to $T_{K}$ with significant
deviations setting in above this temperature scale. This is not surprising given the fact that 
the thermopower is a highly sensitive probe of the particle-hole asymmetry in the spectral density. 

For the thermal conductance, $K_{\rm e}(T)$, we expect from the Wiedemann-Franz law, 
$K_{\rm e}/T \sim G(T)$, to see a scaling in $K_{\rm e}(T)/T$ similar to that in $G(T)/G(0)$. 
This is confirmed in Fig.~\ref{figure7}b which shows $K_{\rm e}(T)/\alpha T$ 
versus $T/T_{K}^{\theta}$ for several gate voltages in the Kondo regime, where 
$\alpha$ is defined by
\begin{equation}
\alpha = \lim_{T\rightarrow 0}\frac{K_{\rm e}(T)}{T}=\frac{\pi^{2}k_{B}^{2}A(0,0)}{3},
\label{alpha}
\end{equation}
and $T_{K}^{\theta}$ is a Kondo scale defined by
\begin{equation}
\frac{K_{\rm e}(T=T_{K}^{\theta})}{T_{K}^{\theta}}=\frac{\alpha}{2}.
\label{tk-theta}
\end{equation}
We see that, for $U/\Gamma=8$, $K_{\rm e}(T)/\alpha T=g(T/T_{K}^{\theta})$ is a universal
function of $T/T_{K}^{\theta}$ for temperatures extending up to at least $100T_{K}^{\theta}$, 
just as $G(T)/G(0)=f(T/T_{K})$ is a universal function of $T/T_{K}$ for temperatures 
extending up to at least $100T_{K}$. Increasing $U/\Gamma$, and thereby reducing $T_{K}/\Gamma$
allows universality to extend to still higher temperatures. Suppressing 
charge fluctuations, e.g. by working within a Kondo model, allows these  
universal scaling functions to be defined for all temperatures.
Note also, that although these universal functions $f$ and $g$ 
have a similar functional dependence on $T/T_{K}$ and $T/T_{K}^{\theta}$ respectively, 
they are shifted relative to one another on an absolute temperature scale. The difference 
between $g$ and $f$ for temperatures around $T=T_{K}^{\theta}\approx T_{K}/1.9$ 
accounts for the violation in 
the Wiedemann-Franz law on this scale, as noted previously (see inset to Fig.~\ref{figure6}g).
The Wiedemann-Franz law is only satisfied exactly in the Fermi liquid regime $T\ll T_{K}$.
One can collapse $G(T)$ onto $K_{\rm e}(T)/T$ by scaling the temperature axis of the former by 
$T_{K}^{\theta}/T_{K}$. In the universal regime $T\ll \Gamma$, small deviations between 
$g(t)$ and $f(t)$ arise for $t\lesssim 1$ and $t\gtrsim 1$. 

For dilute magnetic impurities, our conclusions for scaling in the Kondo regime 
are essentially the same as those above for quantum dots 
(see also Ref.~\onlinecite{costi.94}).

\section{Gate voltage dependence of transport properties}
\label{gate-dependence}

\begin{figure*}[t]
\includegraphics[scale=0.4]{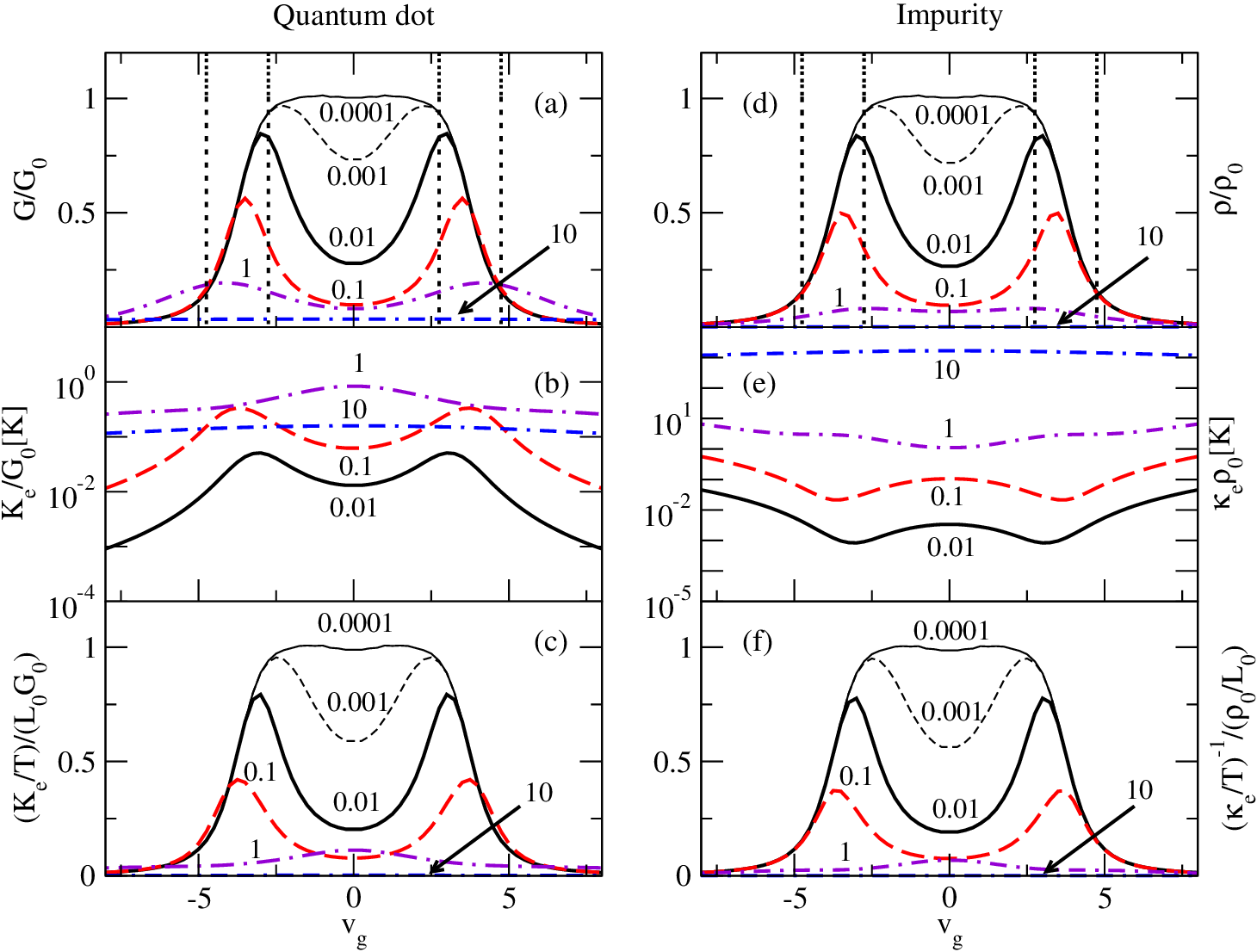}
\caption{ Gate voltage dependence of electrical and thermal 
transport properties of quantum dots (left panels (a-c)) and magnetic impurities 
(right panels (d-f)) for $U/\Gamma=8$ and at several representative temperatures $T/\Gamma$ 
(labeling the individual curves in the figures). Vertical dashed lines in the top panels 
delineate the Kondo regime from mixed valence and empty orbital regimes.
(a): The normalized conductance $G(T)/G_{0}$ ( (d): the normalized resistivity $\rho(T)/\rho_{0}$ 
for the impurity case), and, (b), the rescaled thermal conductance 
$K_{\rm e}(T)/G_{0}$ ( (e): the rescaled thermal conductivity $\kappa_{\rm e}(T)\rho_{0}$ for the 
impurity case). The bottom panel, (c), shows the dimensionless ratio
$(K_{\rm e}/T)/(L_{0}G_{0})$ for the quantum dot (panel (f) shows the
analogous dimensionless ratio for magnetic impurities 
$(\kappa_{\rm e}/T)^{-1}/(\rho_{0}/L_{0})$). Deviations from $G(T)/G_{0}$ 
in (a) reflects deviations from the Wiedemann-Franz law 
(similarly, deviations of $(\kappa_{\rm e}/T)^{-1}/(\rho_{0}/L_{0})$
in (f) from $\rho(T)/\rho_{0}$ in (d) indicate deviations from the 
Wiedemann-Franz law for the impurity case). For clarity not all 
temperatures are shown in (b) and (e).
} \label{figure8}
\end{figure*}

\begin{figure}[t]
\includegraphics[scale=0.325]{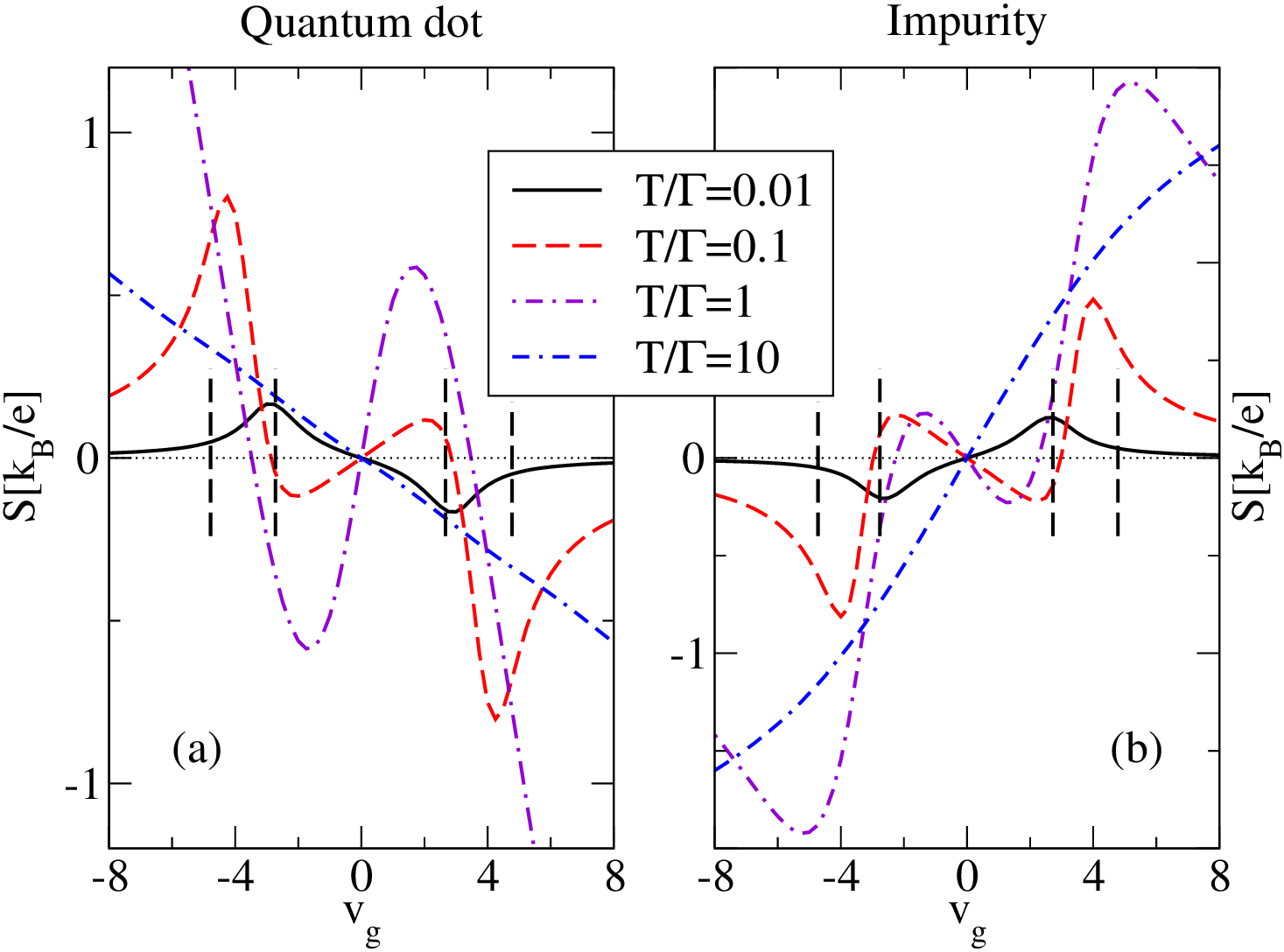}
\caption{Gate voltage dependence of the thermopower for the strong
correlation case $U/\Gamma=8$ 
at several characteristic temperatures $T/\Gamma$ for, (a), the quantum dot case, 
and, (b), the impurity case. The vertical dashed lines delineate the Kondo regime (around
${\rm v}_{g}=0$) from the surrounding mixed valence and empty orbital regimes.
} \label{figure9}
\end{figure}

The gate voltage dependence of transport through  
quantum dots and magnetic impurities is shown for several representative temperatures 
in Fig.~\ref{figure8} for the electrical and thermal transport and in 
Fig.~\ref{figure9} for the thermopower.
For magnetic impurities, different ${\rm v}_{g}$ should be understood as 
corresponding to changes in the local level position $\varepsilon_{d}$ 
relative to the Fermi level, as invoked by the application of pressure
(either chemical via doping or hydrostatic). In some rare earth systems\cite{zlatic.05}
the application of pressure has been shown to tune the magnetic impurities from
the Kondo to the mixed valence and empty orbital regimes. Throughout this section we denote by
$T_{1}$ and $T_{2}$ the minimum temperatures of $T_{1}({\rm v}_g)$ and $T_{2}({\rm v}_g)$ 
in the limit ${\rm v}_g\rightarrow 0$ (see Fig.\ref{figure4} and Table~\ref{tableTK}).
We show that the different behavior in the gate voltage dependence of the
thermopower, at fixed temperature $T$, can be classified in terms of the relative 
value of $T$ to $T_{1}$ and $T_{2}$. 

\subsection{Gate voltage dependence of $G$ and $K_{\rm e}$}
The gate voltage dependence of the electrical 
and thermal conductance of quantum dots is shown in  Fig.~\ref{figure8}a-c. 
The former exhibits, for $T\gg T_{K}$, Coulomb blockade 
peaks at $\varepsilon_d = \pm U/2\Gamma$ with a suppression of the conductance in the
mid-valley region around ${\rm v}_g=0$. 
On decreasing the temperature, the Kondo effect becomes operative resulting in an enhancement
of the conductance in the region between the Coulomb blockade peaks (see Fig.~\ref{figure8}a). 
This picture is well known. At $T\gg T_{K}$, the thermal conductance of quantum dots also exhibits 
Coulomb blockade peaks\cite{zianni.07,tsaousidou.07}, not directly evident in the plot of $K_{\rm e}$ 
versus gate voltage (Fig.~\ref{figure8}b), where only weak signatures of these
are discernible. They become clearer in the gate voltage dependence of $K_{\rm e}/T$ which by
the Wiedemann-Franz law (see Sec.~\ref{wf-lorenz}) is proportional to the electrical conductance $G(T)$, 
as seen in Fig.~\ref{figure8}c. Differences between
Fig.~\ref{figure8}a and Fig.~\ref{figure8}c indicate the degree of 
deviation from the Wiedemann-Franz law. These deviations are largest at $T\gg \Gamma$ for
all gate voltages, as previously observed in Fig.~\ref{figure6}g-i. 

The same observations, using a different terminology, can be made for the case of 
magnetic impurities in Fig.~\ref{figure8}d-f: 
the valence fluctuation peaks at $T\gg \Gamma$ are now seen in the resistivity of 
mixed valence impurities, whereas Kondo impurities (${\rm v}_g\approx 0$) 
have a small resistivity at $T\gg \Gamma$ and a large unitary resistivity at $T=0$. The behavior
of the thermal conductivity $\kappa_{\rm e}$ is similarly understood in terms of the Wiedemann-Franz law
$\kappa_{\rm e}(T)\sim \sigma T$ with $\sigma = 1/\rho$, as seen by comparing 
Fig.~\ref{figure8}d and Fig.~\ref{figure8}f.

\subsection{Gate voltage dependence of $S$}
The gate voltage dependence of the thermopower of quantum dots is shown in Fig.~\ref{figure9}a. 
The particle-hole symmetry about ${\rm v}_{g}=0$ (see Sec.~\ref{model})
implies $S_{-{\rm v}_g}=-S_{{\rm v}_g}$ at
all temperatures, so we only discuss ${\rm v}_g>0$. We focus mainly on the Kondo regime, 
${\rm v}_g<{\rm v}_g^{c}$, and discuss the remaining gate voltages by reference to Fig.~\ref{figure3}. 
There are three main types of behavior, characterized by the following temperatures, $T$, relative
to $T_{1}$ and $T_{2}$: (i), $T<T_{1}$, as exemplified by $T=0.01\Gamma$, 
(ii), $T_{1}<T<T_{2}$ as exemplified by $T=0.1\Gamma$ and $T=\Gamma$, and, 
(iii), $T>T_{2}$, as exemplified by $T=10\Gamma$.
In case (i), the Kondo resonance is asymmetric about the Fermi level \cite{costi.94},
lying slightly above it for ${\rm v}_g>0$. The slope of the spectral density at $\omega=0$ is
positive, resulting by Eq.~(\ref{sommerfeld-exp}) in a negative thermopower, 
as observed for $T=0.01\Gamma$. The same holds, at still lower temperatures,
$T\ll T_{K}$, where Fermi liquid theory \cite{costi.94} gives the explicit expression  
(\ref{thermopower-fl}).
Case (ii), $T_{1}<T<T_{2}$, is the most interesting for quantum dots, for several reasons: 
first, this temperature range is experimentally accessible since for $U/\Gamma =8$, we have
$T_{1}=0.044\Gamma$ and $T_{2}=3.04\Gamma$. Second, there is an overall sign change in $S({\rm v}_g)$,
relative to case (i), for a finite range of gate voltages (see Fig.~\ref{figure9}a). Third, 
a further sign change occurs at finite ${\rm v}_{g}>0$, and, fourth, the thermopower is large enough for a 
significant range of gate voltages to enable its measurement. The sign change at a finite gate 
voltage occurs when $S(T_{1}<T<T_{2})$ as a function of ${\rm v}_g$ 
in Fig.~\ref{figure3}d reaches the value zero and becomes negative. 

For gate voltages outside
the Kondo regime, the thermopower, as a function of gate voltage, $S({\rm v}_g)$, either approaches
zero at ${\rm v}_g\gg 1$ (as happens for $T=0.1$ in Fig.~\ref{figure9}a) or does not saturate 
for ${\rm v}_g\gg 1$ (e.g. for $T=\Gamma$ in Fig.~\ref{figure9}a). In terms of 
Fig.~\ref{figure3}f (see the arrows), the former occurs for temperatures
to the left of the minimum in $S$ in  Fig.~\ref{figure3}f, and the latter occurs for the opposite
case. The latter case is half-way to case (iii), $T>T_{2}\gg \Gamma$, which exhibits a thermopower
approximately linear in gate voltage, with no sign change at any ${\rm v}_g>0$. This
is similar to the ``sawtooth'' behavior of $S({\rm v}_g)$ found for
multi-level quantum dots weakly coupled to leads at $T\gg \Gamma$ in 
Ref.~\onlinecite{beenakker.92,staring.93}(for related experimental work see 
Ref.~\onlinecite{vanHouten.92,molenkamp.92,dzurak.93,cobden.93,dzurak.97,moeller.98}). 
The behavior of the thermopower in multi-level open quantum dots has also been
investigated \cite{andreev.01,turek.02,nguyen.10}. 

The same classification
(i)-(iii), as for quantum dots, can be used to explain the local level dependence of 
the thermopower of magnetic impurities shown in Fig.~\ref{figure9}b. 

\section{Conclusions and discussion}
\label{conclusions}
In this paper we investigated the thermoelectric properties of strongly 
correlated quantum dots, described by the single level Anderson impurity 
model connected to two conduction electron leads. For this purpose, we 
used Wilson's NRG method and calculated the 
local Green's function and transport properties by using the full 
density matrix approach \cite{fdm.07}. Since this approach builds into
the density matrix all excitations obtained in the NRG approach, it is 
particularly well suited to finite temperature transport calculations,
allowing us, for example, to investigate also the high temperature
asymptotics of transport properties.

For strong correlations and in the Kondo regime, the thermopower 
exhibits two sign changes, at temperatures $T_{1}({\rm v}_g)$ and $T_{2}({\rm v}_g)$ 
with $T_{1}< T_{2}$. We found that $T_{1}> T_{p}({\rm v}_g)\approx T_{K}({\rm v}_g)$, 
where $T_{p}({\rm v}_g)$ is the position of the Kondo induced peak in the thermopower, 
$T_{K}({\rm v}_g)$ is the Kondo scale, and $T_{2}= O(\Gamma)$.
The loci of  $T_{1}({\rm v}_g)$ and $T_{2}({\rm v}_g)$ merge at a critical
gate voltage ${\rm v}_g={\rm v}_g^{c}(U/\Gamma)$, beyond which no sign change occurs.
We determined ${\rm v}_g^{c}$ for different $U/\Gamma$ finding that ${\rm v}_g^{c}$
coincides, in each case, with entry into the mixed valence regime. No sign change is 
found outside the Kondo regime or for weak correlations, $U/\Gamma \lesssim 1$.
Thus, a sign change in $S(T)$ at {\em finite} ${\rm v}_g$ is a particularly sensitive signature 
of strong correlations and Kondo physics. This effect could be measurable
in quantum dots, as it manifests itself in an overall sign change in $S$ for a finite range
of gate voltages on increasing temperature $T$ from below $T_{1}({\rm v}_{g}\rightarrow 0)$
to values in the range $T_{1}<T<T_{2} = O(\Gamma)$, which is an accessible 
range since $T_{1}>T_{K}$.

The results for quantum dots were compared also 
to the relevant transport coefficients of dilute magnetic impurities 
in non-magnetic metals: the electronic contribution, $\kappa_{\rm e}$, to the
thermal conductivity, the thermopower, $S$, and the impurity contribution to the 
electrical resistivity, $\rho$. As regards the temperature dependence of the respective
transport quantities, we find, in the mixed valence and empty orbital regimes, 
two peaks in $K_{\rm e}(T)$ as compared to a single peak in 
$\kappa_{\rm e}(T)$. Similarly, $G(T)$ exhibits a finite temperature peak on entering the mixed
valence regime, whereas such a pronounced peak is absent in $\rho(T)$, even far into
the empty orbital regime. As for quantum dots, we find that the low temperature Kondo peak position
in the thermopower of magnetic impurities scales with $T_{K}$. 
We compared and contrasted the  figure of merit,
power factor and the extent of violation of the Wiedemann-Franz law in quantum
dots and dilute magnetic impurities, finding enhanced figures of merit at 
temperatures where the Wiedemann-Franz law is strongly violated.
Finally, we clarified the extent of scaling, as a function of $T/T_{K}$, in the 
thermopower and thermal conductance of quantum dots in the Kondo regime.

\begin{figure}[t]
\includegraphics[scale=0.325]{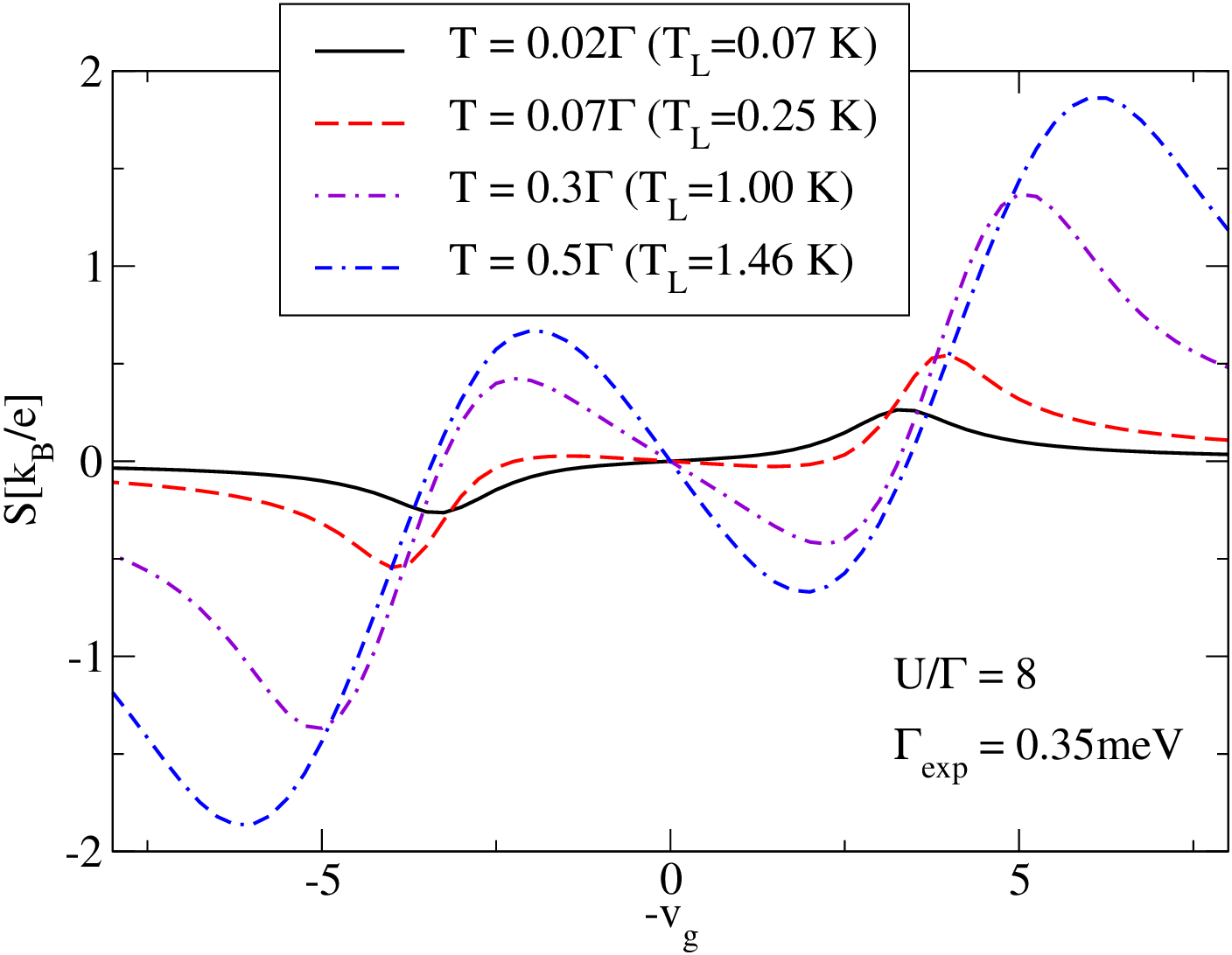}
\caption{Gate voltage dependence of the thermopower for $U/\Gamma=8$ 
at temperatures $T/\Gamma$ corresponding to the experimental ones $T_{L}$ from 
Ref.~\onlinecite{scheibner.05} using $\Gamma_{\rm exp}=0.35 {\rm\, meV}$. Since the experimental gate
voltage $V_{E}\sim -{\rm v}_{g}$ we plotted $S$ versus $-{\rm v}_{g}$ to facilitate
comparing with the experimental data.
} \label{figure10}
\end{figure}

We comment on a recent experiment in Ref.~\onlinecite{scheibner.05} which we believe shows
evidence of Kondo correlations in the thermopower of a strongly correlated
quantum dot. In this experiment, the thermovoltage across a Kondo correlated quantum
dot is investigated as a function of gate voltage and lattice temperature. This can be
compared to our $S({\rm v}_g)$ in Fig.~\ref{figure9}a. The gate voltage
$V_{E}$ in Ref.~\onlinecite{scheibner.05} is related to our dimensionless gate voltage, ${\rm v}_g$,
via $-eV_{E}\sim \varepsilon_{d} \sim {\rm v}_g$,
i.e. $V_{E}\sim -{\rm v}_g$. Mirror reflecting our results for $S({\rm v}_g)$ in Fig.~\ref{figure9}a 
about ${\rm v}_g=0$ allows a qualitative comparison with the experimental
measurements in Ref.~\onlinecite{scheibner.05}. Using the experimental estimate 
$\Gamma=0.35 {\rm\, meV}$ from Ref.~\onlinecite{scheibner.05}, we can translate 
the four experimental temperatures $T_{L}=0.07 {\rm\,K}, 0.25 {\rm\,K}, 
1.00 {\rm\,K}$ and $T_{L}=1.46 {\rm\,K}$ at which the thermopower was measured into our theoretical 
temperatures in units of $\Gamma$. We assume strong Coulomb correlations on 
the dot $U/\Gamma=8$ and show the results for the thermopower in Fig.~\ref{figure10}. 
From Table~\ref{tableTK}, the lowest experimental temperature corresponds to $T\ll T_{1}$, the next
lowest temperature ($T=1.00 {\rm\,K}$) lies close to $T=T_{1}$, where the thermopower changes sign
in the Kondo regime, and the highest two temperatures lie between $T_{1}$ and $T_{2}$.
The lowest temperature measured, $T=0.07 {\rm\,K}$, indeed shows a positive thermopower above mid-valley, 
in agreement with our results for $T<T_{1}$. Upon increasing the temperature, the experiment 
shows a sign change of the thermovoltage for a finite range of gate voltages 
(relative to mid-valley), which is 
consistent with our prediction of such a sign change in the Kondo regime 
for $T_{2}>T>T_{1}$. The onset, with increasing temperature, of an additional oscillation in
$S({\rm v}_g)$ about ${\rm v}_g=0$ in the experiments is therefore consistent with our results.
The experimental data deviates from our calculated thermopower in the mixed valence and empty orbital 
range of gate voltages ${\rm v}_{g}\gg 1$, with the theoretical results showing a much larger 
thermopower in this region of gate voltages. These deviations are expected at ${\rm v}_{g}\gg 1$ 
since additional levels present in real quantum dots, but absent in our model, start 
being populated. This significantly influences transport through the quantum dot.
Qualitatively, however, we are able to interpret these experiments on the thermopower
of Kondo correlated quantum dots for gate voltages ${\rm v}_{g}\approx 0$. 
For a more quantitative comparison to theory, further investigations are needed. 

Calculations for single quantum dots and dilute magnetic impurities, 
are a starting point for dealing with a finite density of quantum dots, 
such as self-assembled quantum dots, or for a finite concentration
of magnetic or mixed valence impurities in bulk (e.g. for Tl impurities
in PbTe in Ref.~\onlinecite{matusiak.09}). The transport properties of
such systems, modeled by a random distribution of Anderson impurities, 
will be determined by (\ref{imp1}-\ref{imp3}) subject to the charge neutrality 
condition $n_{i}n_{d}+n_{c}=n$, where $n_{d}$ is the occupancy 
of the dot (impurity), $n_{c}$ is the occupancy of the relevant conduction band 
and $n$ is the total electron filling. Coupled with material specific electronic 
structure information and the effects of phonons, such calculations, will be 
important for understanding the potential of materials such as self-assembled 
quantum dots or PbTe$_{1-x}$Tl$_{x}$ systems for thermoelectric applications. 

\begin{acknowledgments}
Financial support from the Forschungszentrum J\"{u}lich (V.Z.) and supercomputing 
time from the John von Neumann Institute for Computing (J\"ulich) is gratefully 
acknowledged. We thank J. von Delft and A. Weichselbaum for useful comments and 
discussions.
\end{acknowledgments}
\appendix
\section{Reduction of two-channel Anderson model to a single-channel Anderson model}
\label{reduction}
The reduction of the single-level two-lead Anderson model (\ref{qdot-two-leads}) 
for a quantum dot, to a single-channel
model is, in general, approximate, but as we show here, the approximation is 
very good (or even exact). One notices first, that the $d$-state of the quantum dot 
in (\ref{qdot-two-leads}) 
only couples to the even combination $t_{L}c_{Lk\sigma}+t_{R}c_{Rk\sigma}$
of the lead electron states. By using the following canonical transformation
\begin{eqnarray}
t\; a_{ek\sigma} &=& t_{L}c_{Lk\sigma}+t_{R}c_{Rk\sigma}\\
t\; a_{ok\sigma} &=& t_{L}c_{Rk\sigma}-t_{R}c_{Lk\sigma},
\end{eqnarray}
noting that normalization of even/odd states implies $t^{2}=t_{L}^{2}+t_{R}^{2}$, 
we can rewrite (\ref{qdot-two-leads}) in terms of even ($e$) and odd ($o$) lead states,
as follows
\begin{eqnarray}
H &=& \sum_{k\sigma}\epsilon_{ek\sigma}a_{ek\sigma}^{\dagger}a_{ek\sigma} 
+ \sum_{\sigma}\varepsilon_{d}\,d_{\sigma}^{\dagger}d_{\sigma} + 
U n_{d\uparrow}n_{d\downarrow}\nonumber\\
& + &t\sum_{k\sigma}(a_{ek\sigma}^{\dagger}d_{\sigma}+h.c.) + H_{o} + H_{\rm pot}.
\label{qdot-one-lead}
\end{eqnarray}
Here, $\varepsilon_{ek\sigma}=(\varepsilon_{Lk\sigma}t_{L}^{2}
+\varepsilon_{Rk\sigma}t_{R}^{2})/t^{2}$, 
$H_{o}=\sum_{k\sigma}\epsilon_{ok\sigma}a_{ok\sigma}^{\dagger}a_{ok\sigma}$
is the Hamiltonian for the odd lead electrons with
$\varepsilon_{ok\sigma}=(\varepsilon_{Lk\sigma}t_{R}^{2}+\varepsilon_{Rk\sigma}t_{L}^{2})
/t^{2}$, and 
$H_{\rm pot}=\sum_{k\sigma}U_{k}^{eo}(a^{\dagger}_{ek\sigma}a_{ok\sigma}+h.c.)$ is
a potential scattering term between even and odd lead electrons. Hence, the 
odd lead electrons do not couple to the dot directly, but only indirectly via 
the potential scattering term. The magnitude of this is given by
$U_{k}^{eo}=(\varepsilon_{Lk\sigma}-\varepsilon_{Rk\sigma})
t_{L}t_{R}/t^{2}$, which is vanishingly small at 
low energies. Moreover, it vanishes identically for degenerate leads 
$\varepsilon_{Lk\sigma}=\varepsilon_{Rk\sigma}$. 
The calculations we report in this work, using the single channel Anderson model
\begin{eqnarray}
H &=& \sum_{k\sigma}\epsilon_{ek\sigma}a_{ek\sigma}^{\dagger}a_{ek\sigma} 
+ \sum_{\sigma}\varepsilon_{d}\,d_{\sigma}^{\dagger}d_{\sigma} + 
U n_{d\uparrow}n_{d\downarrow}\nonumber\\
& + &t\sum_{k\sigma}(a_{ek\sigma}^{\dagger}d_{\sigma}+h.c.),
\label{siam-appendix}
\end{eqnarray}
are therefore a very good 
approximation, even in general, to those obtained from the two-lead model 
(\ref{qdot-two-leads}) and identical to those for the case
$\varepsilon_{Lk\sigma}=\varepsilon_{Rk\sigma}$. 
Since $t^{2}=t_{L}^{2}+t_{R}^{2}$, the hybridization strength 
$\tilde{\Gamma} = \pi N_{F} t^{2}$ of the single channel model 
is seen to be the relevant single-particle broadening, $\tilde{\Gamma}_{L}+\tilde{\Gamma}_{R}$, of
the two-lead model (\ref{qdot-two-leads}). In this paper we follow the convention in
the quantum dot community of using the full-width at half-maximum, $\Gamma=2\tilde{\Gamma}$,
as the unit of energy. Finally, we note, that a reduction to a single-channel model is, in general, 
not possible for multi-level or double quantum dots attached to two leads 
\cite{sakano.07,nakanishi.07,cho.05}.

\section{Results for moderate and weak correlations}
\label{extra-results}
\begin{figure}[t]
\includegraphics[scale=0.35]{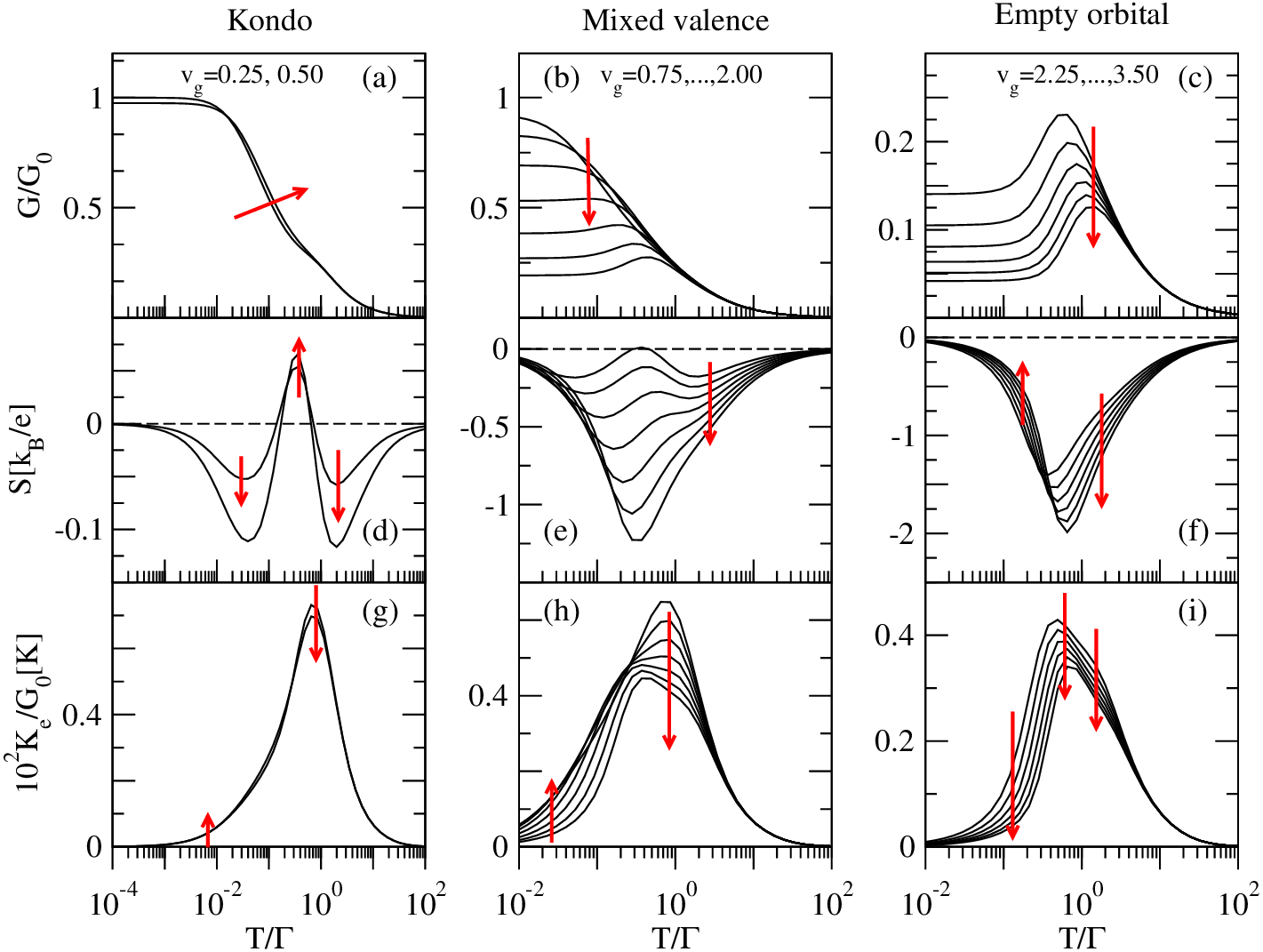}
\caption{  Temperature dependence of, (a-c), the normalized electrical conductance
$G/G_{0}$, (d-f), the thermopower, $S$, and, (g-i), the normalized thermal 
conductance, $K_{\rm e}/G_{0}$, multiplied by a factor $10^2$ for clarity of 
presentation , as function of $T/\Gamma$, in the moderately correlated regime 
$U/\Gamma=3$ and a range of dimensionless
gate voltages, ${\rm v}_g=(\varepsilon_{d}+U/2)/\Gamma>0$, in the Kondo 
(first column), mixed valence (second column) and empty orbital 
(third column) regimes. The range of ${\rm v}_g$ is indicated in the top
panels for each regime and the increment used was $0.25$. Arrows indicate 
the evolution of the transport quantities with increasing ${\rm v}_g$.
} \label{figure11}
\end{figure}

Fig.~\ref{figure11} shows results for a moderately correlated quantum dot,
$U/\Gamma=3$ exhibiting the same trends as those found for the strongly
correlated case $U/\Gamma=8$ (Fig.~\ref{figure3}).
\begin{figure}[t]
\includegraphics[scale=0.275]{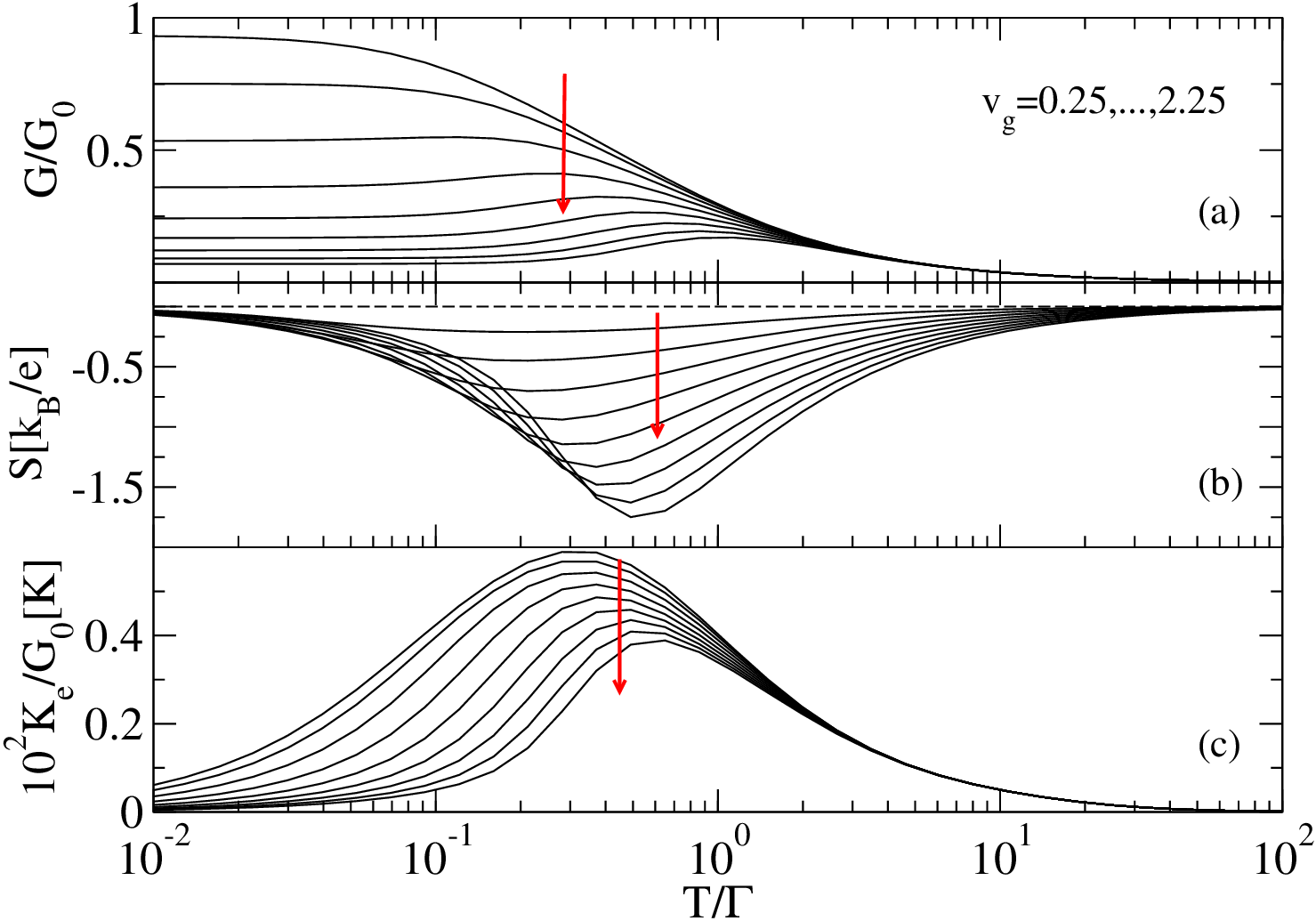}
\caption{  Temperature dependence of, (a),the normalized electrical conductance,
$G/G_{0}$, (b), the thermopower, $S$, and, (c), the normalized thermal conductance, 
$K_{\rm e}/G_{0}$, multiplied by a factor $10^2$ for clarity of presentation, as 
function of $T/\Gamma$, in the weakly correlated regime $U/\Gamma=1$ 
and a range of dimensionless gate voltages, ${\rm v}_g=(\varepsilon_{d}+U/2)/\Gamma$.
The range of ${\rm v}_g$ is indicated in the top
panel and the increment used was $0.25$. Arrows indicate 
the evolution of the transport quantities with increasing ${\rm v}_g$.} \label{figure12}
\end{figure}
For completeness, we show an example of transport through a weakly correlated
quantum dot with $U/\Gamma=1$ in Fig.~\ref{figure12}. In this case, the thermopower
remains negative for all gate voltages ${\rm v}_g>0$  (Fig.~\ref{figure12}b). 
Similarly, the thermal conductance exhibits only a single peak for all gate voltages
(Fig.~\ref{figure12}c). 

\section{Green's functions within the FDM approach}
\label{full-density-matrix}
In this appendix we give an alternative derivation of the finite temperature 
Green's function within the FDM approach of Weichselbaum and
von Delft \cite{fdm.07}. A concise derivation, implementing arbitrary
abelian symmetries, has also been given in Ref.~\onlinecite{toth.08}. We consider a general
fermionic retarded Green's function 
\begin{eqnarray}
G_{AB}(t)&=&-i\theta(t)\langle[A(t),B]_{+}\rangle\nonumber\\
&=&-i\theta(t) {\rm Tr}\left[\rho(A(t)B + BA(t))\right],\nonumber
\end{eqnarray} 
where $A,B$ are fermionic operators, e.g. for
the d-level Green's function of our quantum dot $A=d_{\sigma}$ and $B=d_{\sigma}^{\dagger}$.
The trace is evaluated for an appropriate density matrix $\rho$ by using the complete
set of states introduced by Anders and Schiller \cite{anders.05}. These consist of 
the set of states $|lem\rangle=|lm\rangle|e\rangle$ obtained from the eliminated 
eigenstates, $|lm\rangle$, of $H_{m}$, and the degrees of freedom,
denoted collectively by $e$, of the sites $i=m+1,\dots,N$,
where $N$ is the longest chain diagonalized. The retained low energy states 
of $H_{m}$ are denoted $|km\rangle$, and $|kem\rangle=|km\rangle|e\rangle$ 
extends these to the Hilbert space of $H_{N}$ by the additional environment
degrees of freedom $e$ of the sites $i=m+1,\dots,N$.
The eigenstates (retained and eliminated), $|p=(k,e)m\rangle$, and 
eigenvalues, $E_{p=(k,l)}^{m}$, of $H_{m}$ satisfy $H_{m}|pm\rangle=E_{p}^{m}|pm\rangle$. 
Completeness of the states $|lem\rangle$ is expressed by \cite{anders.05}
\begin{eqnarray}
1&=&\sum_{m'=m_{0}+1}^{N}\sum_{le}|lem'\rangle\langle lem'|
\label{completeness}
\end{eqnarray}
where $m_{0}$ is the last iteration for which all states are retained. For iterations
$m>m_{0}$, the set of states $|p=(k,l)m\rangle$ consists of both retained ($k$) and 
eliminated ($l$) states. The following decomposition of (\ref{completeness}) is useful
\cite{anders.05}:
\begin{eqnarray}
1&=&\sum_{m'=m_{0}+1}^{N}\sum_{le}|lem'\rangle\langle lem'| = 1^{+}_{m} + 1^{-}_{m}\\
1^{-}_{m} &=&\sum_{m'=m_{0}+1}^{m}\sum_{le}|lem'\rangle\langle lem'|\\
1^{+}_{m} &=&\sum_{m'=m+1}^{N}\sum_{le}|lem'\rangle\langle lem'|\nonumber\\
&=&\sum_{ke}|kem\rangle\langle kem|
\label{useful-identity}
\end{eqnarray}  
where the last equation follows from the fact that the Hilbert space of retained states
at iteration $m$ (supplemented by the degrees of freedom $e$ for sites $m'=m+1,\dots,N$)
spans the same Hilbert space as all eliminated states from all subsequent iterations.
By using the decomposition of unity (\ref{completeness}) twice within the trace 
in the expression for $G_{AB}(t)$, the following Lehmann representation can be found for
this Green's function\cite{peters.06}
\begin{widetext}
  \begin{eqnarray}
    G_{AB}(t) &=& G_{AB}^{i} + G_{AB}^{ii} + G_{AB}^{iii},\nonumber\\
    G_{AB}^{i} &=& -i\theta(t)\sum_{m=m_{0}+1}^{N}\sum_{le,l'e'}\left[
    e^{i(E_{l}^{m}-E_{l'}^{m})t}\langle lem|A|l'e'm\rangle
    \langle l'e'm|B\rho|lem\rangle
+e^{i(E_{l}^{m}-E_{l'}^{m})t}\langle lem|A|l'e'm\rangle
    \langle l'e'm|\rho B|lem\rangle\right]\nonumber\\
    G_{AB}^{ii} &=& -i\theta(t)\sum_{m=m_{0}+1}^{N-1}\sum_{le,ke'}\left[e^{i(E_{l}^{m}-E_{k}^{m})t}
    \langle ke'm|B\rho|lem\rangle\langle lem|A|ke'm\rangle
+e^{i(E_{k}^{m}-E_{l}^{m})t}
    \langle lem|\rho B|ke'm\rangle\langle ke'm|A|lem\rangle\right]\nonumber\\
    G_{AB}^{iii} &=&-i\theta(t)\sum_{m=m_{0}+1}^{N-1}\sum_{lem,ke'}\left[e^{i(E_{k}^{m}-E_{l}^{m})t}
    \langle lem|B\rho|ke'm\rangle\langle ke'm|A|lem\rangle
+e^{i(E_{l}^{m}-E_{k}^{m})t}
    \langle ke'm|\rho B|lem\rangle\langle lem|A|ke'm\rangle\right]\nonumber
  \end{eqnarray}
\end{widetext}
where the double sum over $m,m'$ (coming from two applications of (\ref{completeness}))
is decomposed into contributions $m'=m$ (first term), $m'>m$ (second term) and 
$m'<m$ (third term). In the last two terms, use has also been
made of (\ref{useful-identity}). In the time evolution $e^{iHt}|pem\rangle, p=(k,l)$, we have
made use of the NRG approximation $H\approx H_{m}$, so that 
$e^{iHt}|pem\rangle \approx e^{itE_{p}^{m}}|pem\rangle$.
Peters et al. \cite{peters.06} evaluated the above expression for the
Green's function by using an approximate density matrix $\rho_{N}$, 
defined by the eliminated states of the longest chain diagonalized, i.e.
\begin{equation}
\rho_{N} = 
\frac{1}{Z_{N}(T_{N})}\sum_{l}|lN\rangle e^{-\beta_{N} E_{l}^{N}}\langle lN|,
\label{rho-end-of-chain}
\end{equation}
where $\beta_{N}=1/k_{B}T_{N}$ (or $T_{N}$) is chosen appropriately \cite{kww.80}
to ensure that $Z_{N}(T_{N})$ is a good approximation to the partition function of the infinite
system at temperature $T=T_{N}$. Note also, that since $N$ is the last iteration,
all states in the above expression are considered as eliminated states in order that 
(\ref{completeness}) be satisfied. This procedure can be
repeated for each chain length $m=N, N-1,\dots,m_{0}+1$, using a
density matrix 
\begin{equation}
\rho_{m}=\frac{1}{Z_{m}(T_{m})}\sum_{l}|lm\rangle
e^{-\beta_{m}E_{l}^{m}}\langle lm|,
\end{equation}
to obtain shell Green's functions
$\tilde{G}_{m}(\omega), m=N,N-1,\dots,m_{0}+1$ defined at a corresponding set of temperatures
$T_{m}$ (or $\beta_{m}=1/k_{B}T_{m}$) (for clarity we henceforth 
omit the subscript $AB$ for $G_{AB}$). Since these shell Green's functions, $\tilde{G}_{m}$, 
contain only excitations of order the characteristic scale, $\omega_{m}$, of $H_{m}$, or larger, 
the Green's function $\tilde{G}_{m}(\omega)$ can only be evaluated at frequencies 
$\omega > T_{m}$. Information at $\omega\ll T_{m}$ is not available. 
This restriction is overcome by the FDM approach that we now describe.

Weichselbaum and von Delft \cite{fdm.07} evaluated the 
above Green's function by using the FDM of the system made up of 
the complete set of eliminated states from all iterations $m=m_{0}+1,\dots,N$. 
Specifically, the FDM is defined by
\begin{eqnarray}
\rho = \sum_{m=m_{0}+1}^{N}\sum_{le}|lem\rangle \frac{e^{-\beta E_{l}^{m}}}{Z(T)} \langle lem|
\label{fdm}
\end{eqnarray}
where $Z(T)$ is the partition function made up from the complete spectrum, i.e. it contains
all eliminated states from all $H_{m}, m=m_{0}+1,\dots,N$. Consequently, evaluating
the Green's functions by using the above FDM, allows an arbitrary temperature $T$ 
to be used for all frequencies $\omega$, and, in particular, allows accurate calculations to be
carried out at $\omega\ll T$. 

Consider the following density matrix for the m'th shell (defined, however, in the Hilbert space of
$H_N$):
\begin{equation}
\tilde{\rho}_{m} = \sum_{le}|lem\rangle \frac{e^{-\beta E_{l}^{m}}}{\tilde{Z}_{m}} \langle lem|
\end{equation}
Normalization, ${\rm Tr}[\tilde{\rho}_{m}]=1$, implies
\begin{equation}
1 = \sum_{l}\frac{e^{-\beta E_{l}^{m}}}{\tilde{Z}_{m}}4^{N-m}=4^{N-m}\frac{Z_{m}}{\tilde{Z}_{m}}
\end{equation}
where $Z_{m}=\sum_{l}e^{-\beta E_{l}^{m}}$. Then the FDM can be written
as a sum of weighted density matrices for shells $m=m_{0}+1,\dots,N$
\begin{eqnarray}
\rho &=& \sum_{m=m_{0}+1}^{N}w_{m}\tilde{\rho}_{m}\\
w_{m} &=& 4^{N-m}\frac{Z_{m}}{Z}; \sum_{m=m_{0}+1}^{N}w_{m} = 1\label{weights}
\end{eqnarray}
The calculation of the weights $w_{m}$ is outlined in Sec.~\ref{fdm-sub1}.
Substituting $\rho=\sum_{m'}w_{m'}\tilde{\rho}_{m'}$ into the above Lehmann representation for $G(t)$
and Fourier transforming yields $G(\omega)=\sum_{m'}w_{m'}(G^{i}_{m'}(\omega)+G^{ii}_{m'}(\omega)+G^{iii}_{m'}(\omega))$. The first term, $G_{m'}^{i}$, is easily evaluated by using the orthonormality of the
eliminated states $\langle l'e'm'|lem\rangle=\delta_{ll'}\delta_{ee'}\delta_{mm'}$, orthonormality of
environment degrees of freedom in $\langle lem |A|l'e'm'\rangle=\delta_{ee'}A_{ll'}^{m'}$,
with $A_{ll'}^{m'}=\langle lm'|A|l'm'\rangle$ and the trace over the $N-m'$ environment degrees of
freedom in
$$\frac{1}{\tilde{Z}_{m'}}\sum_{e}=\frac{4^{N-m'}}{\tilde{Z}_{m'}}=\frac{1}{Z_{m'}},$$ 
to obtain (for $m'=m_{0}+1,\dots,N$)
\begin{equation}
  G_{m'}^{i}(\omega) =\frac{1}{{Z}_{m'}}\sum_{ll'}A^{m'}_{ll'}B^{m'}_{l'l}
  \frac{(e^{-\beta E^{m'}_{l}} +e^{-\beta E^{m'}_{l'}} )}
       {\omega+E^{m'}_{l}-E^{m'}_{l'}+i\delta}.\nonumber
\end{equation}
The second term, $G_{m'}^{ii}$, is also easily evaluated and results for $m'=m_{0}+1,\dots,N-1$ (the
$N$'th term vanishes, as all states are counted as eliminated states at this iteration)
\begin{eqnarray}
  G_{m'}^{ii}(\omega) &=&\frac{1}{Z_{m'}}\sum_{lk}
  A^{m'}_{lk}B^{m'}_{kl}\frac{e^{-\beta E^{m'}_{l}}}
  {\omega+E^{m'}_{l}-E^{m'}_{k}+i\delta}\nonumber\\
  &+&\frac{1}{{Z}_{m'}}\sum_{kl}A^{m'}_{kl}B^{m'}_{lk}
  \frac{e^{-\beta E^{m'}_{l}}}
       {\omega+E^{m'}_{k}-E^{m'}_{l}+i\delta}\nonumber
\end{eqnarray}
The third term, $G_{m'}^{iii}$, takes the form
\begin{eqnarray}
  G_{m'}^{iii}(\omega) &=&\sum_{lek}A^{m}_{lk}\frac{\langle kem|\tilde{\rho}_{m'}B|lem\rangle}
  {\omega+E^{m}_{l}-E^{m}_{k}+i\delta}\nonumber\\
  &+&\sum_{kle}A^{m}_{kl}\frac{\langle lem|B\tilde{\rho}_{m'}|kem\rangle}
  {\omega+E^{m}_{k}-E^{m}_{l}+i\delta}\nonumber
\end{eqnarray}

Inserting $1=1^{+}_{m}+1^{-}_{m}$ between $\tilde{\rho}_{m'}$ and $B$ in $\langle kem|\tilde{\rho}_{m'}B|lem\rangle$ 
and between  $B$ and $\tilde{\rho}_{m'}$ in $\langle lem|B\tilde{\rho}_{m'}|kem\rangle$
gives 
\begin{eqnarray}
\langle kem|\tilde{\rho}_{m'}B|lem\rangle &=& \langle kem|1^{+}_{m}\tilde{\rho}_{m'}B|lem\rangle\nonumber\\
&+& \langle kem|1^{-}_{m}\tilde{\rho}_{m'}B|lem\rangle\nonumber
\end{eqnarray}
and
\begin{eqnarray}
\langle lem|B\tilde{\rho}_{m'}|kem\rangle &=& \langle lem|B1^{+}_{m}\tilde{\rho}_{m'}|kem\rangle\nonumber\\
&+& \langle lem|B1^{-}_{m}\tilde{\rho}_{m'}|kem\rangle\nonumber
\end{eqnarray}
In Sec.~\ref{fdm-sub2} we show that the second terms in the above expressions vanish, i.e.
\begin{eqnarray}
\langle kem|1^{-}_{m}\tilde{\rho}_{m'}B|lem\rangle&=&0
\label{rho-B}\\
\langle lem|B1^{-}_{m}\tilde{\rho}_{m'}|kem\rangle&=&0.
\label{B-rho}
\end{eqnarray}
On using $1^{+}_{m} =\sum_{k'e'}|k'e'm\rangle\langle k'e'm|$ from 
Eq.(\ref{useful-identity}) the terms involving $1_{m}^{+}$ are evaluated as
\begin{eqnarray}
\langle kem|1^{+}_{m}\tilde{\rho}_{m'}B|lem\rangle &=& 
\sum_{k'e'}\langle kem|\tilde{\rho}_{m'}|k'e'm\rangle\langle k'e'm|B|lem\rangle\nonumber\\
&=&\sum_{k'}\langle kem|\tilde{\rho}_{m'}|k'em\rangle\langle k'em|B|lem\rangle\nonumber\\
&=&\sum_{k'}\langle kem|\tilde{\rho}_{m'}|k'em\rangle B_{k'l}^{m}\nonumber\\
\langle lem|B1^{+}_{m}\tilde{\rho}_{m'}|kem\rangle &=& 
\sum_{k'e'}\langle lem|B|k'e'm\rangle \langle k'e'm|\tilde{\rho}_{m'}|kem\rangle\nonumber\\
&=&\sum_{k'}\langle lem|B|k'em\rangle \langle k'em|\tilde{\rho}_{m'}|kem\rangle\nonumber\\
&=&\sum_{k'}B_{lk'}^{m} \langle k'em|\tilde{\rho}_{m'}|kem\rangle\nonumber
\end{eqnarray}
Note that these expressions are finite only for $m'> m$. Using the definition
of the reduced density matrix\cite{hofstetter.00}  
$$\rho_{\rm red}^{m'\rightarrow m}(k,k')= {\rm Tr_{e}}[\langle kem |\tilde{\rho}_{m'}|k'em\rangle],
$$
we arrive at the following expression for $G_{m'}^{iii}(\omega)$
\begin{eqnarray}
  G_{m'}^{iii}(\omega) &=&\sum_{m=m_{0}+1}^{m'-1}
\sum_{lkk'}A^{m}_{lk}\frac{{\rho}_{\rm red}^{m'\rightarrow m}(k,k')B_{k'l}^{m}}
  {\omega+E^{m}_{l}-E^{m}_{k}+i\delta}\nonumber\\
  &+&\sum_{m=m_{0}+1}^{m'-1}\sum_{lkk'}A^{m}_{kl}\frac{{\rho}_{\rm red}^{m'\rightarrow m}(k',k)B_{lk'}^{m}}
  {\omega+E^{m}_{k}-E^{m}_{l}+i\delta}\nonumber
\end{eqnarray}
Hence, the final expression for $G(\omega)=\sum_{m'=m_{0}+1}^{N}w_{m'}(G_{m'}^{i}(\omega)
+G_{m'}^{ii}(\omega)+G_{m'}^{iii}(\omega))$ is given by
\begin{eqnarray}
  G(\omega) &=&\sum_{m'=m_{0}+1}^{N}\frac{w_{m'}}{{Z}_{m'}}
\sum_{ll'}A^{m'}_{ll'}B^{m'}_{l'l}
  \frac{(e^{-\beta E^{m'}_{l}} +e^{-\beta E^{m'}_{l'}} )}
       {\omega+E^{m'}_{l}-E^{m'}_{l'}+i\delta}\nonumber\\
  &+&\sum_{m'=m_{0}+1}^{N-1}\frac{w_{m'}}{Z_{m'}}\sum_{lk}
  A^{m'}_{lk}B^{m'}_{kl}\frac{e^{-\beta E^{m'}_{l}}}
  {\omega+E^{m'}_{l}-E^{m'}_{k}+i\delta}\nonumber\\
  &+&\sum_{m'=m_{0}+1}^{N-1}\frac{w_{m'}}{{Z}_{m'}}\sum_{kl}A^{m'}_{kl}B^{m'}_{lk}
  \frac{e^{-\beta E^{m'}_{l}}}
       {\omega+E^{m'}_{k}-E^{m'}_{l}+i\delta}\nonumber\\
  &+&\sum_{m=m_{0}+1}^{N-1}
\sum_{lkk'}A^{m}_{lk}\frac{{\rm R}_{\rm red}^{m}(k,k')B_{k'l}^{m}}
  {\omega+E^{m}_{l}-E^{m}_{k}+i\delta}\nonumber\\
  &+&\sum_{m=m_{0}+1}^{N-1}\sum_{kk'l}A^{m}_{kl}\frac{{\rm R}_{\rm red}^{m}(k',k)B_{lk'}^{m}}
  {\omega+E^{m}_{k}-E^{m}_{l}+i\delta}\nonumber
\end{eqnarray}
where, in the last two terms we rearranged the summations over $m'$ and $m$ and 
introduced the full reduced density matrix
$$
{\rm R}_{\rm red}^{m}(k,k') = \sum_{m'=m+1}^{N} w_{m'}\rho_{\rm red}^{m'\rightarrow m}(k,k').
$$
Note that the meaning of this quantity is completely analogous to the reduced density matrix 
introduced by Hofstetter in Ref.~\onlinecite{hofstetter.00} except that one obtains reduced
density matrices at iteration $m$ by eliminating environment degrees of freedom $e=e_{m+1}e_{m+2}...e_{N}$
from the FDM (\ref{fdm}) instead of the density matrix for iteration $N$. In addition,
the former
is built from the complete set of eliminated states, as opposed to the retained states of iteration $N$
in the approach of Ref.~\onlinecite{hofstetter.00}.
The above expression for $G(\omega)$ is identical to that in Ref.~\onlinecite{fdm.07}. We have checked that
the sum rule for the spectral function $A_{\sigma}(\omega,T)=-\frac{1}{\pi}{\rm Im}[G_{AB}(\omega)]$
$$
\int_{-\infty}^{+\infty} A_{\sigma}(\omega,T) d\omega = 1
$$
is satisfied exactly (to machine precision) when using the discrete (unbroadened) form
of the spectral function as in Ref.~\onlinecite{fdm.07}.
 
\subsection{Calculation of weights $w_{m}$}
\label{fdm-sub1}
The expression for $w_{m}$ in (\ref{weights}) involves $Z$ which contains eigenvalues
from all iterations $m'=m_{0}+1,\dots,N$. In evaluating these expressions, one should
therefore use the absolute energies for the $E_{l}^{m}$. Since, in practice, the 
iterative diagonalization of the Hamiltonian $H_m$ involves subtraction of groundstate 
energies and rescaling at each $m$ (see Ref.~\onlinecite{kww.80}), one has to keep track 
of the subtracted groundstate energies and return to the actual
physical energies relative to a common absolute energy reference in evaluating $w_{m}$ 
and $Z$. We take this absoute energy reference to be the ground state energy of the last 
Wilson iteration $N$. Thus, if $E_{GS}^{m}$ is the true groundstate energy of $H_{m}$, we use 
$E_{l}^{m}\rightarrow E_{l}^{m}+E_{GS}^{m}$, $Z_{m}\rightarrow e^{-\beta E_{GS}^{m}}Z_{m}$
in evaluating $w_{m}/Z_{m}$ and $Z$
\begin{eqnarray}
\frac{w_{m}}{Z_{m}} &=& \frac{4^{N-m}e^{-\beta E_{GS}^{m}}}{\sum_{m'=m_{0}+1}^{N}4^{N-m'}
e^{-\beta E_{GS}^{m'}}Z_{m'}}\nonumber\\
Z &=& \sum_{m'=m_{0}+1}^{N}4^{N-m'}e^{-\beta E_{GS}^{m'}}Z_{m'}\nonumber
\end{eqnarray}
\subsection{Proof of Eq.~(\ref{rho-B}) and Eq.~(\ref{B-rho})}
\label{fdm-sub2}
Using the expression for $\tilde{\rho}_{m'}$ we easily find that
\begin{eqnarray}
1^{-}_{m}\tilde{\rho}_{m'}&=&\sum_{m''=m_{0}+1}^{m}\sum_{le}|lem''\rangle\langle lem''|\times\nonumber\\
&&\sum_{l'e'}|l'e'm'\rangle\frac{e^{-\beta E_{l'}^{m'}}}{\tilde{Z}_{m'}}\langle l'e'm'|\nonumber\\
&=&\sum_{m''=m_{0}+1}^{m}\delta_{m''m'}\sum_{le}|lem'\rangle\frac{e^{-\beta E_{l'}^{m'}}}{\tilde{Z}_{m'}}  
\langle lem'|\nonumber\\
&=& \left\{ \begin{array}{ll}
         \tilde{\rho}_{m'} & \mbox{if $m' \leq m$};\\
        0 & \mbox{if $m'>m $}.\end{array} \right.
\end{eqnarray}
Hence $\langle kem|1^{-}_{m}\tilde{\rho}_{m'}B|lem\rangle$ in Eq.~(\ref{rho-B}) involves matrix elements
of the form $\langle kem|l'e'm'\rangle$ for $m'\leq m$, which vanish, since all retained state
at iteration $m$ have no overlap with eliminated states at iterations $m'\leq m$ (i.e., eliminated
states of previous iterations are not used to obtain retained states of later iterations). The
same arguments can be used to prove Eq.~(\ref{B-rho}).

\section{Thermal conductance and thermopower of quantum dots}
\label{transport-derivations}
For completeness, we outline here the derivation of thermoelectric transport through a 
strongly interacting quantum dot\cite{dong.02,kim.02}. The electrical, $I_{L}$,
and heat current, $I^{Q}_{L}$, from the left lead to the quantum dot can be 
expressed in terms of the particle number $N_{L}=\sum_{k\sigma}c_{Lk\sigma}^{\dagger}c_{Lk\sigma}$ 
and energy $H_{L}=\sum_{k\sigma}\epsilon_{Lk\sigma}c_{Lk\sigma}^{\dagger}c_{Lk\sigma}$ 
of the left lead, via
\begin{eqnarray}
I_{L}&=&-e\dot{N}_{L}=-\frac{e}{i\hbar}[N_{L},H]\\
I^{Q}_{L}&=&\dot{H}_{L}-\mu_{L}\dot{N}_{L}=\frac{1}{i\hbar}[H_{L}-\mu_{L}N_{L},H], 
\end{eqnarray}
where $H$ is the Hamiltonian (\ref{qdot-two-leads}). In terms
of the lesser Green's function's 
$G^{<}_{d\sigma,kL\sigma}(t,t')=i\langle c^{\dagger}_{kL\sigma}(t') d_{\sigma}(t)\rangle$ and
$G^{<}_{kL\sigma,d\sigma}(t,t')=i\langle d^{\dagger}_{\sigma}(t') c_{kL\sigma}(t)\rangle = 
-(G^{<}_{d\sigma,kL\sigma})^{*}$, the above currents are given by
\begin{eqnarray}
I_{L}&=& \frac{2e}{\hbar}Re\left[\sum_{k\sigma} t_{L}G^{<}_{d\sigma,kL\sigma}(t,t)\right],\\
I^{Q}_{L}&=&-\frac{2}{\hbar}Re\left[\sum_{k\sigma} t_{L}(\varepsilon_{Lk\sigma}-\mu_{L})
G^{<}_{d\sigma,kL\sigma}(t,t)\right].
\end{eqnarray}
The lesser Green's function $G^{<}_{d\sigma,kL\sigma}(t,t')$ can be expressed via equations of
motion solely in terms of Green's functions of the dot and the non-interacting Green's function
for the left lead. After some lengthy algebra \cite{kim.02,jauho.94}, 
one finds the following expressions for the
currents in terms of the retarded, $G_{d\sigma}^{r}=G_{d\sigma}(\omega+i\delta)$, advanced, 
$G_{d\sigma}^{a}=G_{d\sigma}(\omega-i\delta)$ 
and lesser Green's function, $G_{d\sigma}^{<}(\omega)$ of the dot:
\begin{eqnarray}
I_{L}&=&\frac{ie}{\hbar}\sum_{\sigma}\int d\omega\tilde{\Gamma}_{L} \left[(G_{d\sigma}^{<}(\omega)\right. \nonumber\\
&+&\left. f_{L}(\omega)(G_{d\sigma}^{r}-G_{d\sigma}^{a})\right],\\
I^{Q}_{L}&=&-\frac{i}{\hbar}\sum_{\sigma}\int d\omega(\omega-\mu_{L})
\tilde{\Gamma}_{L}\left[ G_{d\sigma}^{<}(\omega)\right. \nonumber\\
&+&\left. f_{L}(\omega)(G_{d\sigma}^{r}-G_{d\sigma}^{a})\right],
\end{eqnarray}
where $f_{L}(\omega)=(1+e^{-(\omega-\mu_{L})/(k_{B}T_{L})})^{-1}$ is the Fermi 
function of the left lead and $\tilde{\Gamma}_{L}=\pi N_{F}t_{L}^{2}$ is the hybridization strength 
of the dot to the left lead as defined in Sec.~\ref{model}. 
By using current conservation $I_{L}=-I_{R}$, one can eliminate the lesser Green's function from the above
expressions to arrive at the final expressions used in this paper
\begin{eqnarray}
I_{L}&=&\frac{e}{\hbar}\sum_{\sigma}\int d\omega (f_{L}-f_{R}){\cal T}_{d\sigma}(\omega),\\
I^{Q}_{L}&=&-\frac{i}{\hbar}\sum_{\sigma}\int d\omega (\omega-\mu_{L})
(f_{L}-f_{R}){\cal T}_{d\sigma}(\omega).
\end{eqnarray}
The quantity ${\cal T}_{d\sigma}(\omega)$ acts as a transmission function and is given by
\begin{eqnarray}
{\cal T}_{d\sigma}(\omega) &=& 2i\frac{\tilde{\Gamma}_{L}\tilde{\Gamma}_{R}}{\tilde{\Gamma}_{L}+\tilde{\Gamma}_{R}}(G_{d\sigma}^{r}-G_{d\sigma}^{a})
\end{eqnarray}
The electric and heat currents are expanded to linear order in $\delta T = T_{L}-T_{R}$ and
$\delta V = V_{L}-V_{R}$
\begin{eqnarray}
\begin{pmatrix} I_L \cr I^{Q}_L \end{pmatrix} 
 &=& \begin{pmatrix} L_{11} & L_{12} \cr L_{21} & L_{22} \end{pmatrix}
    \begin{pmatrix} \delta V \cr \delta T \end{pmatrix},
\end{eqnarray} 
defining, thereby, the transport coefficients $L_{i,j},i,j=1,2$. In terms of the latter, the
transport properties are given by
\begin{eqnarray}
G(T)&=&\lim_{\delta V\rightarrow 0}I_{L}/\delta V|_{\delta T = 0} = L_{11}
\label{transport-quantities-lij-G}\\
S(T)&=&-\lim_{\delta T\rightarrow 0}{\delta V/\delta T}{|}_{I_{L}=0}= L_{12}/L_{11}
\label{transport-quantities-lij-S}\\
K_{\rm e}(T)&=&-\lim_{\delta T\rightarrow 0}I^{Q}/\delta T{|}_{I_{L}=0}\nonumber\\
&=&L_{12}L_{21}/L_{11} - L_{22}.
\label{transport-quantities-lij-K}
\end{eqnarray} 
Finally, the $L_{ij}$ are simply expressed in terms of the following transport integrals
\begin{eqnarray}
I_n (T) &=& \frac{2}{h} \int d\epsilon ~ \epsilon^n T(\epsilon) 
   \left[ -\frac{\partial f }{ \partial \epsilon} \right], 
\end{eqnarray}
via $L_{11}=e^{2}I_{0}, L_{21}=-eI_{1}/T$ and $L_{22}=I_{2}/T$. Substituting these values for $L_{ij}$ into
(\ref{transport-quantities-lij-G}-\ref{transport-quantities-lij-K}) results in the 
expressions (\ref{transport-expressions-G}-\ref{transport-expressions-K}) given in the text.
%
%


\begin{thebibliography}{99}
\bibitem{specific-heat-note} The quantity $\gamma\sim 1/T_{K}$ is the linear coefficient
  of specific heat of the quantum dot. The specific heat, $C(T)$, of a quantum dot 
  is difficult to measure, but theoretically it is easily calculated from the
  free energy \cite{kww.80}. The linear coefficient of specific heat, $\gamma$, is then obtained
  via $\gamma =\lim_{T\rightarrow 0}C(T)/T$.
\end{thebibliography}
\end{document}